\documentclass[letterpaper,fleqn]{cas-sc}
\usepackage{graphicx}
\usepackage{amsmath,amssymb,amsfonts,mathrsfs,bm}
\usepackage[numbers,sort]{natbib}
\usepackage{subcaption}
\usepackage{siunitx}

\begin{document}
\let\WriteBookmarks\relax
\def\floatpagepagefraction{1}
\def\textpagefraction{.001}

\title[mode = title]{Two-fluid modeling of heat transfer in flows of dense suspensions}
\shorttitle{Two-fluid modeling of heat transfer in flows of dense suspensions}

\author[1]{Pranay P.\ Nagrani}[orcid=0000-0003-4568-9318]
\credit{Formal analysis, Validation, Data Curation, Visualization, Writing}
\author[2,3]{Federico Municchi}[orcid=0000-0002-5105-6173]
\credit{Software, Methodology, Formal analysis, Supervision, Writing}
\author[1]{Amy M.\ Marconnet}[orcid=0000-0001-7506-2888]
\credit{Conceptualization, Methodology, Project administration, Supervision, Writing}

\author[1]{Ivan C.\ Christov}[orcid=0000-0001-8531-0531]
\cormark[1]
\ead{christov@purdue.edu}
\credit{Conceptualization, Methodology, Funding acquisition, Formal analysis, Project administration, Supervision, Writing}
\shortauthors{P.~P.~Nagrani \textit{et al.}}

\address[1]{School of Mechanical Engineering, Purdue University, West Lafayette, Indiana 47907, USA}
\address[2]{Faculty of Engineering, University of Nottingham, University Park, Nottingham NG7 2RD, UK}
\address[3]{School of Mathematical Sciences, University of Nottingham, University Park, Nottingham, NG7 2RD, UK}

\cortext[cor1]{Corresponding author}

\begin{abstract}
We develop a two-fluid model (TFM) for heat transfer in dense non-Brownian suspensions. Specifically, we propose closure relations for the inter-phase heat transfer coefficient and the thermal diffusivity of the particle phase based on calibration against experimental data. The model is then employed to simulate non-isothermal flow in an annular Couette cell. We find that, when the shear rate is controlled by the rotation of the inner cylinder, both the shear and thermal gradients are responsible for particle migration. Within the TFM framework, we identify the origin and functional form of a ``thermo-rheological'' migration force that rationalizes our observations. Furthermore, we apply our model to flow in eccentric Couette cells. Our simulations reveal that the system's heat transfer coefficient is affected by both the classic shear-induced migration of particles and the newly identified thermo-rheological migration effect.
\end{abstract}


\begin{keywords}
Two-fluid model \sep
Interphase heat transfer \sep
Concentrated suspension \sep
Shear-induced migration \sep
Thermo-rheological flux
\end{keywords}

\maketitle

\section{Introduction}
\label{sec:Intro}
Particles in sheared suspension flows tend to migrate from regions of large shear rate to regions of small shear rate. This phenomenon known as particle migration can affect mixing and heat transfer \cite{GuaMor}. Understanding particle migration is important for a broad range of applications including geothermal energy recovery \citep{Wu2016HeatFluid,Wu2017HeatApplications}, hydraulic fracturing \cite{Barbati2016ComplexFracturing,Dontsov2019LubricationJamming}, hemodynamics \citep{Segre1961RadialSuspensions}, microfluidics \cite{DiCarlo2007ContinuousMicrochannels,vanDinther2012SuspensionGradients}, electronics cooling \citep{Colangelo2017CoolingContribution,Dbouk2019AParticles}, and food processing \citep{Lareo1997TheReview,Lin2003ShearConcentrates}. These applications have driven significant research on the topic of fluidized particles and particulate suspensions. However, thermal transport in such ``complex fluids'' appears to be less well understood, especially in the case of dense suspensions. Often approximate models with limited generality and limited validation are applied to estimate effective material properties of a particle-fluid mixture. To fill this gap in the literature, we develop a continuum two-fluid model (TFM) for computational analysis of the interaction between heat transfer and particle migration in the flow of dense suspensions. Specifically, by extending the dense-suspension TFM from \citep{MNC19} to incorporate coupled heat transfer within each and \emph{between} the phases, we wish to provide a general and extensible computational framework with which one can predictively simulate (upon proper calibration performed herein) the interaction between shear-rate gradients and thermal gradients in dense suspensions, across a variety of flow scenarios. 

After briefly introducing the background and state-of-the-art of dense suspension modeling in Section~\ref{sec:SotA}, we discuss the governing equations pertaining to the TFM for modeling flows of dense non-Brownian suspensions, including their rheology in Section~\ref{sec:Math_Flow} and their thermophysical properties in Section~\ref{sec:Math_HT}. We describe the heat transfer closure relation that we propose to capture coupled thermal-particle migration phenomena via the TFM. In Section~\ref{sec:thermo-rheol}, we provide a conceptual argument, based on the temperature-dependent fluid properties of how shear and thermal gradients couple and yield migration forces on the particulate phase. Then, in Section~\ref{sec:Math_CFD}, we summarize the numerical methods used to implement the TFM in OpenFOAM\textsuperscript{\textregistered}, along with the schemes used, and the criteria employed to obtain accurate results. We calibrate our computational model against the heat pulse experiments of \citet{Metzger2013HeatDiffusion} in Section~\ref{sec:Val_CC} to obtain the fitting parameters in the inter-phase heat transfer closure. Then, in Section~\ref{sec:Val_EC}, we (qualitatively) validate the flow solver (no fitting parameters in this isothermal case) against data from \citet{Subia1998ModellingEquation} for an eccentric Couette cell. Next, to understand the interplay between thermal and shear gradients, we study particle migration subject to two thermal boundary conditions (BCs) in a Couette cell (Section~\ref{sec:CC}) --- one BC in which the thermal and shear gradients are in the same direction, and another BC in which they are in opposite directions. We also extend the discussion to eccentric Couette cells (Section~\ref{sec::EccentricDisc}) and show that the geometric eccentricity can be used to tune the thermal performance of the suspension flow. Finally, in Section~\ref{sec:Conclusion}, we summarize our findings and briefly discuss potential avenues for future work.

\section{Background and problem statement}
\label{sec:SotA}

\subsection{Shear-induced particle migration in flows of dense suspensions}
The migration of particles from regions of high shear rate to regions of low shear occurs at low particle Reynolds number (\textit{i.e.}, the regime of negligible particle inertia) and large particle Pecl\'et number  (\textit{i.e.}, the non-Brownian regime) \cite[p.~2]{Morris2020TowardSuspensions}. Historical overviews of this phenomenon can be found elsewhere \citep{Morris2020TowardSuspensions,Denn2014RheologySuspensions}. A standard setup for experimentally characterizing suspension flows is the Couette cell (\textit{i.e.}, the gap between two, usually concentric, rotating cylinders).  Past experimental studies sought to explain, with varying levels of fidelity, the irreversible particle migration phenomenon \cite{Pine2005ChaosSuspensions,Metzger2010IrreversibilitySuspensions}, including its dependence on the shear rate, the particle volume fraction, and the particle size. 

The simplest model of shear-induced particle migration is a phenomenological transport process, in which scalar diffusive fluxes (in principle, dependent on the gradients of shear rate and particle concentration) are posited \citep{Leighton1987MeasurementSpheres,Phillips1992AMigration,Wu2017HeatCylinders}. The model resulting from this approach is termed the \emph{diffusive flux model} (DFM). The DFM is  phenomenological and requires empirical calibration of (\textit{i}) a collision flux (\textit{i.e.}, migration of particles to a region of low collisional frequency between particles \cite{Leighton1987MeasurementSpheres}) and (\textit{ii}) a viscosity flux (\textit{i.e.}, migration caused caused by viscosity gradients  \cite{Leighton1987MeasurementSpheres}). The simplicity of the DFM is appealing, and it yields itself to straightforward computational studies. However, as \citet[p.~212]{Denn2014RheologySuspensions} note, the DFM  ``gives results that conflict with several experiments in other simple-shear geometries'' beyond the Couette cell.

In order to obtain deeper insight into particle migration, the \emph{suspension balance model} (SBM) was introduced by \citet{Nott1994Pressure-drivenTheory} and further refined in recent years \cite{Lhuillier2009MigrationSuspensions,Nott2011TheRevisited}. Rather than treating particle migration simply as an extra diffusive flux in the particle transport equation, the SBM (in its most common implementation \cite{GuaMor}) involves solving the suspension's momentum equation, a particle transport equation, and an energy conservation equation for the suspension. For a pressure-driven flow,  \citet{Nott1994Pressure-drivenTheory} used both particle-resolved Stokesian dynamics simulations and the SBM to explain how particles, starting with a uniform particle volume fraction, migrate to the center of the channel (where the shear rate is low) in inhomogeneous flow due to a force arising from the ``average particle pressure.''

In general, normal stresses arise in flows in curvilinear geometries, and they must be accounted for to understand particle migration. To this end,  \citet{Morris1999CurvilinearStresses} reformulated the SBM \citep{Nott1994Pressure-drivenTheory} to capture the anisotropy of the normal stresses. Their results were further confirmed by the more recent experiments by \citet{Dbouk2013NormalSuspensions}. Consequently, modeling of shear-induced particle migration in flows beyond the canonical unidirectional configuration (such as contraction-expansions and cavity flows \citep{Miller2009SuspensionGeometries}, as well as flows in concentric \citep{Dbouk2013Shear-inducedTensor} and eccentric  \citep{Mirbod2016Two-dimensionalModel} Couette cells) became possible. These SBM formulation \citep{Morris1999CurvilinearStresses,Dbouk2013Shear-inducedTensor} highlight that the divergence of the particle stress tensor, which arises from the stress inhomogeneity, is the driving force for particle migration.

In contrast to the traditional implementation of the SBM, which lumps the fluid and particles into a single phase, \emph{two-fluid models} (TFMs) \cite{DP99} solve the governing (mass and momentum conservation) equations for the particle phase and for the fluid phase separately. A TFM, therefore, does not make the equilibrium assumption between the two phases, which leads to the traditional form of the SBM. Thus, it is expected that a TFM can be applied to a wider class of flows. \citet{Buyevich1996ParticleFlow} used this approach to address the coupled effect of Brownian and shear-induced migration in concentrated suspensions. More recently, TFMs have been applied to simulate proppant transport in hydraulic fracturing applications \citep{Shiozawa2016SimulationSimulator,Dontsov2019LubricationJamming} due to their ability to capture the governing physics in different flow regimes. Meanwhile, \citet{MNC19} implemented and benchmarked a TFM for the numerical simulation of dense suspension flows (including shear-induced migration) in OpenFOAM\textsuperscript{\textregistered}. They showed good agreement with the previous works on particle migration, suggesting general curvilinear flows of dense particulate suspensions can be accurately simulated with a TFM using the proper rheological closures that account for particle stress anisotropy. Here, we build upon this approach.

\subsection{Heat transfer in flows of dense suspensions}
While particle migration has been studied extensively from a fluid mechanics perspective, the effect of thermal gradients and heat transfer on particle migration has not received as much attention. Yet, when suspensions are sheared, experiments have demonstrated that the effective suspension thermal conductivity is enhanced \citep{Ahuja1975AugmentationResults,Ahuja1975AugmentationData,Sohn1981MicroconvectiveExperiment,Shin2000ThermalFields,NiaziArdekani2018NumericalParticles}. In addition to the shear rate, the thermal conductivity of suspensions also depends on other properties such as the particle volume fraction, the particle size, and the particles' thermal diffusivity \citep{Ahuja1975AugmentationData,Ahuja1975AugmentationResults,Shin2000ThermalFields}. Recent research (\textit{e.g.}, \cite{Metzger2013HeatDiffusion,Wu2016HeatFluid,Dbouk2018HeatSimulation}) has focused on modeling this enhancement of thermal transport using effective properties. Our goal is to increase the modeling fidelity by using a TFM.

Early work by \citet{Sohn1981MicroconvectiveExperiment} demonstrated that shearing suspensions induces particle motion in the mixture, which leads to convection that enhances thermal transport. \citet{Shin2000ThermalFields} showed that the suspension's homogenized thermal conductivity increases with the shear rate, though their experiments appear to be in an inertial flow regime. More recently, \citet{Metzger2013HeatDiffusion} performed a heat-pulse experiment, in which they heated the inner cylinder of a Couette cell until steady state was achieved, and then let it cool while observing the temperature decay with and without shearing of the suspension in the gap (via rotation of one cylinder of the cell). A faster temperature decay was observed when the suspension was sheared, suggesting a shear-induced improvement of thermal transport. \citet{Metzger2013HeatDiffusion} developed a closure relation for the effective (homogenized) thermal diffusivity of the suspension as a function of particle volume fraction $\phi$ and the thermal P\'eclet number $Pe_\mathrm{th}$, by drawing upon functional forms motivated in earlier work on shear-enhanced diffusion \cite{Zydney1988AugmentedSuspension,Wang2009HydrodynamicSpheres}.

A limited number of computational studies have addressed the interplay between shear- and thermal-driven particle migration in suspensions. Most recently, \citet{Wu2017HeatCylinders} and  \citet{Dbouk2018HeatSimulation} modified the DFM and SBM, respectively, to account for thermal transport in dense suspension flows. \citet{Wu2017HeatCylinders} then employed their DFM to show a discernible effect of a temperature gradient across a Couette cell's gap (both concentric and eccentric) on the radial particle distribution profile. In a similar vein, \citet{Kang2021OnsetSuspensions} performed a computational DFM study of the onset of thermal convection in suspensions. \citet{Dbouk2018HeatSimulation}, on the other hand, incorporated a conjugate heat transfer model and the closure relation of \citet{Metzger2013HeatDiffusion} into the SBM. He quantified the enhancement of thermal performance due to a suspension, compared to a clear liquid, for forced convection through a rectangular channel. On this basis, \citet{Dbouk2019AParticles} suggested exploiting particle migration effects to improve heat transfer in applications related to CPU cooling. Recently, this model has been applied to buoyancy-driven flows in immersed granular beds \citep{Dbouk2021ModelingBeds}. In our work, gravity plays a minor role since we focus on shear-driven flows.

While these recent works begin to demonstrate the importance of the coupled particle migration and thermal transport, DFMs and the typical form of the SBM do not capture the inter-phase heat transfer between the particle phase and the fluid phase, since they employ a single thermal transport equation, based on effective properties, for the mixture. Specifically, the prior computational models are based on the assumption of thermal equilibrium between the two phases, which we relax in the present work.

\section{Governing equations of the TFM}
\label{sec:Math}

\subsection{Flow and rheology}
\label{sec:Math_Flow}

In this subsection, we summarize the basic equations of the TFM for dense particulate suspensions, as detailed in \cite{MNC19} and \cite{Wang2020ContinuumReview}. Introducing the particle volume fraction field $\phi(\bm{x},t)$, where $\bm{x}$ is the position vector in 3D space and $t$ is time, we write the governing equations for the two phases (`$p$' for particle and `$f$' for fluid) as:
\begin{align}
\frac{\partial }{\partial t} \left( \rho_p \phi \right) + \bm{\nabla} \cdot \left( \rho_p \bm{u}_p \phi \right)
&= 0,\label{eq::continuity_p}\\
\frac{\partial }{\partial t} \left[ \rho_f \left(1-\phi \right)  \right] + \bm{\nabla} \cdot \left[ \rho_f \bm{u}_f \left(1-\phi \right) \right] &= 0,\label{eq::continuity_f}\allowdisplaybreaks\\
\frac{\partial }{\partial t} \left( \rho_p \phi \bm{u}_p \right) + \bm{\nabla} \cdot \left( \rho_p  \phi \bm{u}_p \otimes \bm{u}_p \right)
&=  \bm{\nabla} \cdot \bm{\Sigma}_p  + \phi \rho_p \bm{g} + \bm{f}_d, \label{eq::equilibrium_p}\\
\frac{\partial }{\partial t} \left[ \rho_f \left(1-\phi \right) \bm{u}_f \right] + \bm{\nabla} \cdot \left[ \rho_f  \left(1-\phi \right) \bm{u}_f \otimes \bm{u}_f \right] &= - \ \bm{\nabla} \cdot \left( p\bm{I} -  \bm{\tau}_f \right) - \bm{f}_d + (1-\phi) \rho_f \bm{g}.\label{eq::equilibrium_f}
\end{align}
Equations~\eqref{eq::continuity_p} and \eqref{eq::continuity_f} are the conservation of mass (continuity) equations for the two phases, while Eqs.~\eqref{eq::equilibrium_p} and \eqref{eq::equilibrium_f} are the corresponding conservation of linear momentum equations.

Here, the particle-phase stress tensor $\bm{\Sigma}_p$ is to be modeled, $\bm{\tau}_f$ is the deviatoric stress tensor of the generalized Newtonian fluid phase, and 
\begin{equation}
    \bm{f}_d = K_d  \left( \bm{u}_p - \bm{u}_f \right) + \phi \bm{\nabla} \cdot \left( \bm{\tau}_f - p\bm{I} \right) - \left(1-\phi\right) \boldsymbol{\bm{\nabla}} p_p
\label{eq::drag_force}
\end{equation}
is the inter-phase force, where $K_d$  is the so-called Clift drag coefficient \cite{DiFelice1995HydrodynamicsFluidisation}. Furthermore, $p=p_f+p_p$ is the `shared' pressure, which satisfies the Poisson equation in the case of an incompressible suspension. Importantly, as in \cite{MNC19}, we incorporate the state-of-the art rheological models for dense suspensions via
\begin{equation}\label{eq::Sigma_p}
\bm{\Sigma}_p = 2\mu_p \dot{\bm{S}}_p + \lambda_p\left(\bm{\nabla} \cdot \bm{u}_p \right) \bm{I} + \bm{\Sigma}_s,
\end{equation}
where $\mu_p$ and $\lambda_p$ are the shear and bulk viscosities obtained using the kinetic theory of granular flows \citep{Gidaspow1986HydrodynamicsModeling} and the general expression for the frictional viscosity described in \citep{Municchi2019TwoFluidsNBSuspensionFoam}. For either phase (`$p$' or `$f$'), $\dot{\bm{S}} = \frac{1}{2}[\bm{\nabla} \bm{u} + ( \bm{\nabla} \bm{u})^T ] -  (\bm{\nabla} \cdot \bm{u} ) \bm{I}$ is the deviatoric rate of strain. 

The particulate phase's frictional viscosity $\mu_p$ is expressed, as in \citep{MNC19,Municchi2019TwoFluidsNBSuspensionFoam}, as the product of the fluid viscosity and a function of the particle volume fraction $\eta(\phi)$. Specifically, it takes the form:
\begin{equation}
    \label{eq::mu_fric}
    \mu_p = \mu_f(T_f) \eta \left(\phi \right)\,, \qquad \eta(\phi) = a_{\mu} +b_{\mu}\phi \left(1-\frac{\phi}{\phi_\mathrm{m}}\right)^{-1} + c_{\mu} \left(1-\frac{\phi}{\phi_\mathrm{m}}\right)^{-2}\,, 
\end{equation}
where $\phi_\mathrm{m}$ is the maximum packing fraction  (here, taken to be $0.68$ corresponding to BBC sphere packing), while $a_{\mu}$, $b_{\mu}$, and $c_{\mu}$ are parameters fitted from data in the literature. We recall from \citep{MNC19,Municchi2019TwoFluidsNBSuspensionFoam} that Eq.~\eqref{eq::mu_fric} returns the closures from \citep{Morris1999CurvilinearStresses} and \citep{Maron1956ApplicationParticles}, under appropriate choices of the model parameters. 

The extra contribution $\bm{\Sigma}_s$ in Eq.~\eqref{eq::Sigma_p} is the anisotropic stress, due to the shearing of the particle phase, given by
\begin{equation}\label{eq::Sigma_s}
    \bm{\Sigma}_s = - \mu_f \eta_\mathrm{N} (\phi) \dot{\gamma}_\mathrm{eff} \bm{Q},\qquad \dot{\gamma}_\mathrm{eff} = \left(2 \dot{\bm{S}}_p:\dot{\bm{S}}_p\right)^{1/2} + \dot{\gamma}_\mathrm{NL},
\end{equation}
where $\eta_\mathrm{N}$ is the normal scaled viscosity. The nonlocal shear rate $\dot{\gamma}_\mathrm{NL}$ regularizes the model by accounting for the average stress at the (sub-continuum) particle scale  \cite{Miller2006NormalSuspensions}. Specifically, $\dot{\gamma}_\mathrm{NL}$ ensures that $\dot{\gamma}_\mathrm{eff} \neq 0$, for example, at the centerline of a channel \cite{MNC19}, as a way to overcome the breakdown of the continuum assumption at the particle scale \cite{Nott1994Pressure-drivenTheory,Morris1999CurvilinearStresses,Mills1995RheologyMigration}.

In Eq.~\eqref{eq::Sigma_s}, the extra stress' anisotropy is represented by means of the tensor $\bm{Q}$ \cite{Morris1999CurvilinearStresses}. This anisotropy tensor can be diagonalized by employing a suitable local orthonormal coordinate system based on the particle phase velocity field: $\bm{Q} = \sum_{i=1}^3 \lambda_i\left( \phi \right) \bm{e}_i \otimes \bm{e}_i$, where $\lambda_i \left( \phi \right)$ are the anisotropy weight functions \cite{Morris1999CurvilinearStresses}, and $\bm{e}_i$ are the unit vectors in the direction of the flow ($i=1$), gradient ($i=2$) and vorticity ($i=3$), given by $\bm{e}_1 = \bm{u}_p \big/ |\bm{u}_p|$, $\bm{e}_3 = (\bm{\nabla} \times \bm{u}_p) \big/ |\bm{\nabla} \times \bm{u}_p|$, and $\bm{e}_2 = \bm{e}_1 \times \bm{e}_3$. In DFMs and SBMs, it is not possible to capture the dense suspension's stress anisotropy in this way because of the assumptions made on Eq.~\eqref{eq::equilibrium_p} do not allow for $\bm{u}_p$ to be resolved. The above definition of the unit vectors, introduced in \cite{MNC19}, allows for the straightforward generalization of the model to 3D curvilinear flows.

\subsection{Heat transfer and energy equations}
\label{sec:Math_HT}

To succinctly describe how we incorporate heat transfer within the TFM, it is most convenient to express each phase's energy equation using a mixed formulation with their enthalpies, internal energies, and temperatures:
\begin{align}
    \overbrace{\frac{\partial }{\partial t} \left( \rho_p \phi H_p \right)}^{\text{unsteady}} \;\; +  \;\; \overbrace{\bm{\nabla} \cdot \left( \rho_p \phi H_p \bm{u}_p \right)}^{\text{convection}} \;\;  &= \overbrace{\phi\frac{\partial p}{\partial t}}^{\text{pressure work}} + \;\;  \overbrace{\bm{\nabla}\cdot\left(\rho_p\alpha_p\phi \bm{\nabla} \mathfrak{e}_p\right)}^{\text{conduction}} \;\; - \overbrace{K_h (T_p-T_f),}^{\text{inter-phase heat transfer}} \label{eq::energy_p}\\
    \hspace{-2mm}\frac{\partial }{\partial t} \left[ \rho_f \left(1- \phi \right) H_f \right] + \bm{\nabla} \cdot \left[ \rho_f \left(1-\phi\right) H_f \bm{u}_f \right] &= \left(1-\phi\right)\frac{\partial p}{\partial t} + \bm{\nabla}\cdot\left[\rho_f\alpha_f(1-\phi) \bm{\nabla} \mathfrak{e}_f\right] + K_h (T_p-T_f), \label{eq::energy_f}
\end{align}
where the stagnation (or ``total'') enthalpies $H_{p,f} = \mathfrak{h}_{p,f} + \tfrac{1}{2}|\bm{u}_{p,f}|^2$ have been introduced for convenience from the specific enthalpies $\mathfrak{h}_{p,f}$ of the phases. Note that the internal energy of each phase is $\mathfrak{e}_{p,f} = \mathfrak{h}_{p,f} - p/\rho_{f,p}$. In equations~\eqref{eq::energy_p} and \eqref{eq::energy_f}, $K_h$ is the (volumetric, \si{\watt\per\meter\tothe{3}\per\kelvin}) inter-phase heat transfer coefficient to be modeled; $\alpha_{f}=k_{f}/(\rho_{f}C_{p,f})$ and $\alpha_{p}=k_{p}/(\rho_{p}C_{p,p})$ are the phases' thermal diffusivities, with $k_{p}$, $k_{f}$ and $C_{p,p}$, $C_{p,f}$ being their thermal conductivities and specific heats, respectively; $T_{p}$ and $T_{f}$ are the phases' individual temperature fields. In Eqs.~\eqref{eq::energy_p} and \eqref{eq::energy_f}, we have neglected viscous dissipation (unlike previous work \citep{Dbouk2018HeatSimulation}). This assumption will be justified upon specifying the flow conditions in Section~\ref{sec:RAndD} below. Also, there are no volumetric sources of heat present or work done by gravitational forces. Work done by buoyancy is not considered in the energy equation in present work, as buoyancy was found to have marginal effects in flows similar to those studied herein \citep{Dbouk2018HeatSimulation}. 

To create a predictive TFM for heat transfer in dense suspensions, we must ensure that $K_h$ takes into account the non-uniform shear-induced migration within the particle phase, as well as the particle phase's thermal conductivity relative to the suspending fluid. In the fluidized beds literature \cite{Kuipers1993ComputerBed,Gunn1978TransferBeds}, heat transfer is incorporated into TFMs via a Nusselt correlation: \textit{e.g.}, the Ranz--Marshall \cite{Ranz1952EvaporationI} formula $Nu_{d_p} = 2 + 0.6Re_{d_p}^{1/2}Pr^{1/3}$, where $Re_{d_p} = U_f d_p/\nu_f$ and $Pr = \nu_f/\alpha_f$, then, $K_{h,0} = Nu_{d_p} \,k_f / d_p^2$. Based on this approach, we propose a shear-dependent inter-phase heat transfer coefficient
\begin{equation}
    \label{eq::interPhaseHeatTransferCoeff}
    K_h(\phi,\dot\gamma) = K_{h,0}[1 + \beta\phi(\underbrace{\|\dot{\bm{S}}_p\|d_p^2/\alpha_p}_{\sim Pe_\mathrm{th}})^m],
\end{equation}
where $\beta$ and $m$ are parameters that must be calibrated against experiments (in Section~\ref{sec:Val_CC} below for the heat-pulse experiment from \citep{Metzger2013HeatDiffusion}), $\dot\gamma = \sqrt{2} \|\dot{\bm{S}}_p\|$, and $Pe_\mathrm{th}$ is a particle-based thermal P\'eclet number for the sheared suspension.

Although our expression for $K_h(\phi,\dot\gamma)$ is similar (in functional form) to the mixture models'  $\alpha_\mathrm{eff}(\dot\gamma)$ \citep{Metzger2013HeatDiffusion,Dbouk2018HeatSimulation}, $K_h$ and $\alpha_\mathrm{eff}$ represent fundamentally different physics (see also \citep{Chen2008HeatTheory}). In the TFM, the usual $\alpha_{p}$ and $\alpha_{f}$ take care of conduction within each phase, thus they cannot depend on $\dot\gamma$. Importantly, by writing down separate energy equations for the phases, we shed the assumption of ``microscopic local thermal equilibrium between the solid and fluid phases'' \cite[p.~437]{Dbouk2018HeatSimulation} used in the SBM.

\subsection{The origin of thermo-rheological fluxes}
\label{sec:thermo-rheol}
As we show in Section~\ref{sec:RAndD}, the presence of temperature gradients in a flow results in a net flux of particles which, in most fluids, is oriented in the direction opposite to the heat flux. This phenomenon results from the interplay between the thermal state of the fluid phase and the rheology of the suspended particulate phase. In particular, the dynamic viscosity $\mu_f$ of liquids is generally a decreasing function of temperature \cite[p.~117]{Panton2013IncompressibleFlow}, so that $\mu_f = \mu_f \left(T_f\right)$ and $d \mu_f / d T_f < 0$.

Now, consider the force $\bm{f}_{\Sigma} = \bm{\nabla} \cdot \boldsymbol{\Sigma}_p$ acting on the particulate phase due to the changes in particle stress. Using the chain rule, it is also possible to split $\bm{f}_{\Sigma}$ as
\begin{equation}
     \label{eq::rheo_force_chain_T}
     \bm{f}_{\Sigma} = \underbrace{\mu_f \bm{\nabla} \cdot \widetilde{\bm{\Sigma}}_p}_{\substack{\text{shear-induced} \\ \text{ migration and diffusion}}  } -\;\;  \overbrace{\bm{
     \Sigma}_p \cdot \left( \beta_{\mu}  \bm{\nabla} T_f\right)}^{\text{thermo-rheological}}\,,
\end{equation}
where we defined the scaled particle stress tensor $\widetilde{\bm{\Sigma}}_p = \bm{\Sigma}_p/\mu_f$ and the thermal variation coefficient of the dynamic viscosity:
\begin{equation}
     \label{eq::KTR}
     \beta_{\mu} = \left|\frac{d \left(\ln{\mu_{f}}\right)}{d T_f} \right|\,.
\end{equation}
The second term on the right-hand side of Eq.~\eqref{eq::rheo_force_chain_T} represents the thermo-rheological force that couples the energy equation with the momentum equation. Specifically, Eq.~\eqref{eq::rheo_force_chain_T} shows that there is an additional force oriented in the direction opposite to the projection of the particle stress on the temperature gradient. Hence, if we consider the case in which the particles are pushed in the positive $x$-direction by the shear-induced migration forces, a temperature gradient aligned with the shear rate gradient will result in an opposing flux. It is also clear that no flux can arise solely due to the temperature gradient, but the presence of a shear flow is also required. However, while shear-induced particle migration is driven by the spatial variation of the shear-rate, the thermo-rheological flux should also be observable in uniformly sheared suspensions.

\begin{figure}
    \centering
    \includegraphics{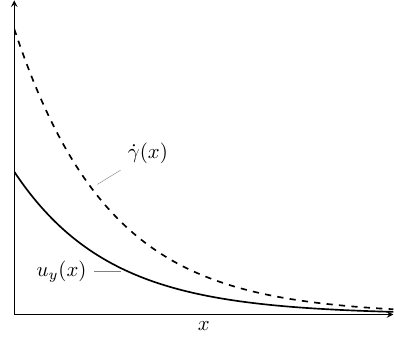}
    \caption{Diagram of a unidirectional shear configuration, which is representative of Couette flow between two cylinders. The continuous line represents the velocity in the $y$-direction and the dashed line represents the shear rate.}
    \label{fig::plot_shear}
\end{figure}

Now, disregarding the terms arising from kinetic theory (which do not play a significant role in the present study due to the low value of the granular temperature and the assumption of a non-Brownian suspension), we can express $\bm{f}_{\Sigma}$, via Eq.~\eqref{eq::Sigma_p}, as:
\begin{equation}
    \label{eq::stress_force_noKin}
    \bm{f}_{\Sigma} = \bm{\nabla} \cdot \left[ 2 \mu_f (T_f) \eta(\phi) \dot{\bm{S}}_p - \mu_f(T_f) \eta_\mathrm{N} (\phi) \dot{\gamma}_\mathrm{eff} \bm{Q} \right] .
\end{equation}
To better illustrate the point, consider a unidirectional flow scenario as depicted schematically in Fig.~\ref{fig::plot_shear}. In this situation, the force is acting in the $x$-direction, normal to the shear, and the suspension is subject to a temperature gradient along $x$. We denote the shear rate as $\dot{s} (x) = \dot{S}_{p,yx} = (1/2) \partial u_{p,y}/\partial x$. Furthermore, we take $\dot{\gamma}_{\mathrm{eff}} = \dot{\gamma} = \sqrt{2}|\dot{s}| = -\sqrt{2}\dot{s} $ as in this case the velocity is always decreasing in the $x$-direction. This configuration is illustrative of the velocity and shear profiles expected in Couette cells. Under these conditions, after some algebra, Eq.~\eqref{eq::stress_force_noKin} becomes:
\begin{equation}
    \label{eq::stress_force_x_final}
    f_{\Sigma,x} =  A (\phi) \left( \mu_f \frac{ \partial \dot{\gamma}}{\partial x} - \dot{\gamma}\left| \frac{d\mu_{f}}{d T_f}\right| \frac{\partial T_f}{ \partial x}  \right) + \mu_f \dot{\gamma} \frac{d A(\phi)}{d \phi} \frac{\partial \phi}{\partial x}\,, \qquad A(\phi) =    -\sqrt{2}\eta   - \eta_\mathrm{N}\lambda_1.
\end{equation}
It is now clear that an additional flux (second term in the parentheses above) opposing the classical migration flux (first term) is induced by the gradient of the shear rate. Equation \eqref{eq::stress_force_x_final} also shows the linear dependence of the thermo-rheological force on the shear rate. Notice that, since $A(\phi)$ is always negative, the thermo-rheological force is oriented in the same direction as the temperature gradient.

In the absence of other forces perpendicular to the flow direction (such as gravity or pressure gradients), the force balance on the particle phase is $f_{\Sigma,x} = 0$. Hence, in order to achieve a perfectly mixed suspension without particle segregation (i.e., $\partial \phi /\partial x$ = 0) the following relation between shear rate and temperature gradient should hold:
\begin{equation}
    \label{eq::balance_no_segregation}
    \frac{\partial \ln{\dot{\gamma}}}{\partial x} = - \beta_{\mu} \frac{\partial T_f}{ \partial x}\,.
\end{equation}
Under the assumption that $\beta_{\mu}$ is not varying significantly, Eq.~\eqref{eq::balance_no_segregation} can be easily integrated for a domain bounded between two walls $x=w_1$ and $x=w_2$, leading to:
\begin{equation}
    \label{eq::gammadot_T_noSeg}
    \ln{\left(\frac{\dot{\gamma}|_{w_1}}{\dot{\gamma}|_{w_2}} \right)} = -\beta_{\mu} \left( T_{f}|_{w_1} - T_{f}|_{w_2} \right)\,.
\end{equation}
Equation \eqref{eq::gammadot_T_noSeg} shows that, if the shear rate is uniform ($\dot{\gamma}|_{w_1} = \dot{\gamma}|_{w_2}$), then only an isothermal suspension can also be homogeneous. We also remark that the integration of the logarithm can be carried out only if $\dot{\gamma} \neq 0$ throughout the domain. This observation also has a physical significance since no thermo-rheological flux can exist where $\dot{\gamma}=0$, while classical migration fluxes might still be present, as they depend on the gradient of $\dot{\gamma}$. Hence, a suspension cannot be homogeneous under such circumstances; the classical migration flux would remain unbalanced at such points. It is worth noting that such regions (or isolated) points where $\dot{\gamma} = 0$ are pathological also in standard shear-induced migration, and a non-local non-zero effective shear rate has to be used instead (recall the discussion following Eq.~\eqref{eq::Sigma_s}). 

Finally, in the more general case in which variations of $\beta_{\mu}$ are not negligible (for example due to large temperature gradients), it is more convenient to integrate the right-hand-side of Eq.~\eqref{eq::balance_no_segregation} with respect to $\mu_f$ rather than $T_f$, leading to:
\begin{equation}
    \label{eq::gammadot_T_noSeg_2}
    \frac{\dot{\gamma}|_{w_1}}{\dot{\gamma}|_{w_2}}  = \frac{\mu_f (T_{f}|_{w_1})}{\mu_f (T_{f}|_{w_2})} \,.
\end{equation}
While Eq.~\eqref{eq::gammadot_T_noSeg_2} is a more complete expression, it possesses the same characteristics as Eq.~\eqref{eq::gammadot_T_noSeg}. However, though conceptually enlightening, both expressions are generally of little practical use as the value of the shear at the walls due to dense suspension flow is generally unknown \emph{a priori} (as there is no analytical solution).

\subsection{Simulation methodology}
\label{sec:Math_CFD}

The governing equations [Eqs.~\eqref{eq::continuity_p}, \eqref{eq::continuity_f}, \eqref{eq::equilibrium_p}, \eqref{eq::equilibrium_f}, \eqref{eq::energy_p}, and \eqref{eq::energy_f}] are solved numerically via the \emph{finite-volume method} (FVM) \cite{Moukalled2016TheMatlab} in a solver implemented in OpenFOAM\textsuperscript{\textregistered} v7. The full description of the numerical approach (based on the earlier algorithm of \citet{Passalacqua2011ImplementationGrids}) can be found in \citep{MNC19}. The FVM discretization ensures that mass is conserved ``automatically'' in all the cell elements that the flow geometry is divided into. A structured mesh was employed for computational modeling of coupled thermal-particle migration in the 2D Couette cell (Fig.~\ref{fig::geoMesh}). A grid independence study was performed (summarized in Appendix~\ref{sec:appendix}) to determine the optimal number of mesh elements needed to obtain accurate results, while simultaneously ensuring reasonable wall-clock time required to complete each simulation.

For the discretization of the transient terms, a second-order backward scheme was employed. Although this scheme, on an orthogonal mesh, is unconditionally stable, this is rarely the case in reality. Deferred correction for high-order schemes, coupling, and nonlinear terms cannot be handled implicitly, introducing a Courant--Friedrichs--Lewy (CFL) number $Co$ constraint \cite{Moukalled2016TheMatlab}, which we enforce to ensure convergence of the simulations. Diffusion terms were discretized using the ``Gauss'' approach in OpenFOAM\textsuperscript{\textregistered}. A second-order linear discretization scheme was employed to obtain higher accuracy for the discretization of diffusion terms. Furthermore, to account for the effect of non-orthogonality of the mesh employed on the discretization of the diffusion terms, a ``corrected" method was used \citep{Moukalled2016TheMatlab}. The linear interpolation scheme from  OpenFOAM\textsuperscript{\textregistered} was chosen to interpolate the diffusion coefficients from the cell faces to the cell centers. The divergence terms were discretized using the ``Gauss'' method of OpenFOAM\textsuperscript{\textregistered}, and an upwind scheme was employed on the convective terms in the equations. The ``cellMDLimited Gauss linear" scheme was used for discretization of the gradient terms. The gradient scheme ensures boundedness of the gradient terms after discretization. In the simulations below, the pressure and velocity residual convergence criteria were set to $10^{-7}$ and $10^{-9}$, respectively. 

The ``PIMPLE'' method, which is a combination of the \emph{pressure-implicit with splitting of operators} (PISO) method and the \emph{semi-implicit method for pressure-linked equations} (SIMPLE) \citep{Moukalled2016TheMatlab} was used to couple the Navier--Stokes and energy equations and obtain converged residuals for the transient simulations. The minimum number of linear solver iteration was always set to 1, so that convergence of simulations was achieved due to convergence of residuals. Moreover, the number of times the entire system of equations was solved was determined by observing the number of iterations required for the residuals to converge during each time step. A dynamic adjustable time step was used in the simulations, which ensured that $\max Co < 0.5$  during the entire simulations. 

\section{Calibration and validation of the TFM}
\label{sec:Val}

\subsection{Heat transfer through a sheared suspension in a concentric Couette cell}
\label{sec:Val_CC}

\begin{figure}
    \centering
    \includegraphics[width=0.9\linewidth]{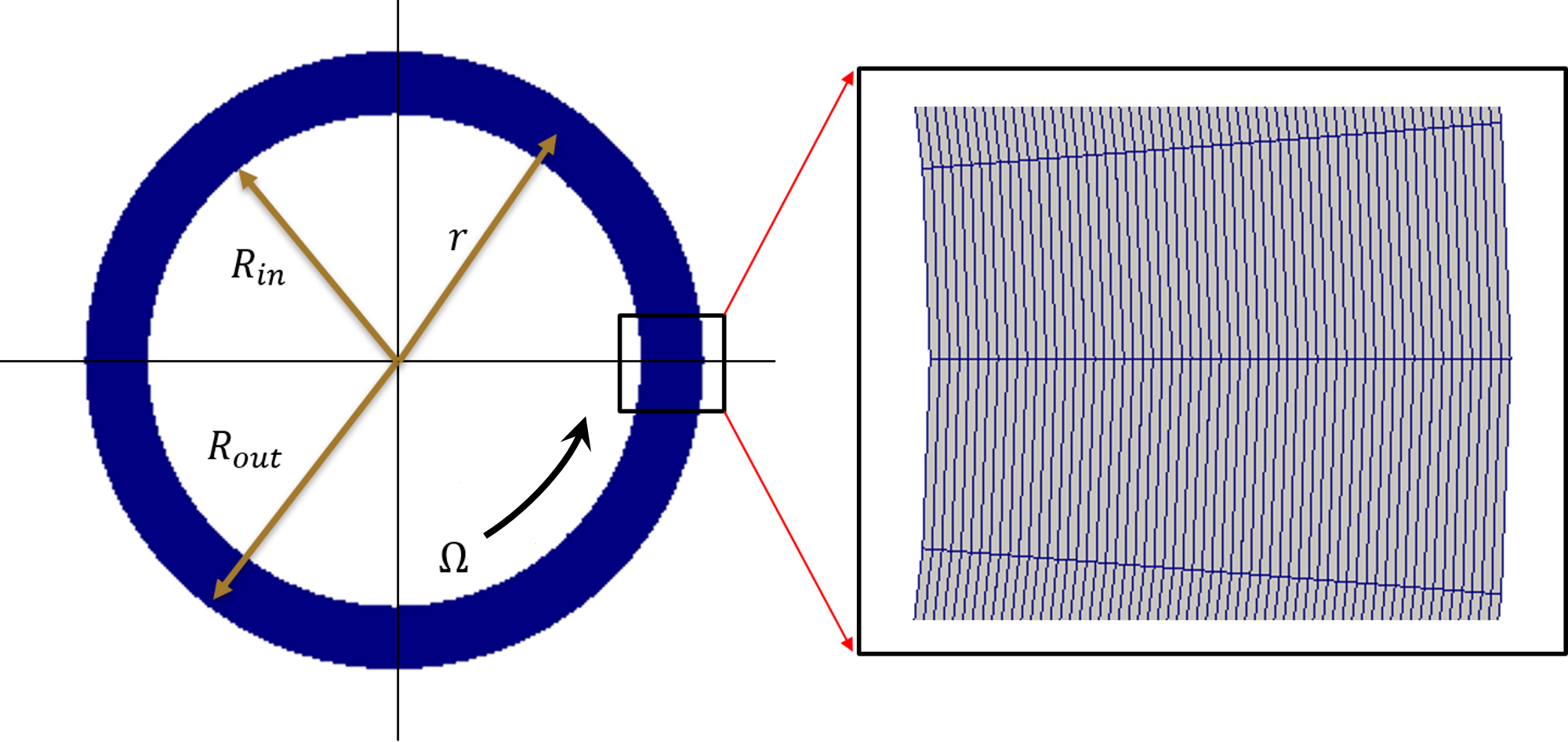}
    \caption{Geometry and structured computational mesh of the annular Couette cell's flow domain.}
    \label{fig::geoMesh}
\end{figure}

\citet{Metzger2013HeatDiffusion} evaluated heat transfer in a sheared suspension of poly-methyl methacrylate (PMMA) particles dispersed within a Newtonian fluid (mixture of Triton X-100, zinc chloride solution and water) in a Couette cell geometry (top view shown in Fig.~\ref{fig::geoMesh}). The particles and fluid had identical thermophysical properties in order to isolate the effect of shear-induced migration on the heat transfer enhancement. We calibrate the parameters in our TFM via their experimental data.

To simulate the shearing of the suspension in a concentric Couette cell with inner and outer radii of $R_\mathrm{in} = 5~\si{\centi\meter}$ and $R_\mathrm{out} = 6.2~\si{\centi\meter}$, respectively, we use a structured mesh (shown in Fig.~\ref{fig::geoMesh}). The inner cylinder rotates at a given rate of $\Omega_\mathrm{in}$, while the outer cylinder is held stationary ($\Omega_\mathrm{out}=0$). For convenience, we book-keep the applied shear rate at steady state via the relation $\dot{\gamma}= (2 \Omega_\mathrm{in} R_\mathrm{out}^2) /(R_\mathrm{out}^2 - R_\mathrm{in}^2)$ at the inner wall, which admittedly only holds true for a clear Newtonian fluid flow. In what follows, this quantity is reported, and the value of $\dot{\gamma}$ is set by varying $\Omega_\mathrm{in}$. The fluid and particle thermophysical properties used are listed in Table~\ref{tab::sus_props}. Consistent with the experiments of \citep{Metzger2013HeatDiffusion}, we use a neutrally buoyant suspension ($\rho_p = \rho_f$) with an initial, uniform bulk particle volume fraction of $\phi(\bm{x},t=0)=\phi_b$ everywhere across the gap. We do not consider the effect of temperature varying thermophysical properties as the maximum temperature difference in the system is only $5~\si{\kelvin}$ in this case.

\begin{table}
    \centering
    \begin{tabular}{l|l|l}
      Property  & Particle (`$p$') & Fluid (`$f$')\\
      \hline
      \hline
      Density $\rho$ (\si{\kilo\gram\per\meter\tothe{3}}) & 1180 & 1180 \\
      \hline
      Dynamic viscosity $\mu$ (\si{\pascal\second}) & -- & 3.0 \\
      \hline
      Thermal conductivity $k$ (\si{\W\per\m\per\kelvin}) & 0.19 & 0.19 \\
      \hline
      Specific heat $C_p$ (\si{\J\per\kg\per\kelvin}) & 1260 & 1260 \\
    \end{tabular}
    \caption{Particle and fluid properties used for the calibration of the TFM against the experimental results of \citet{Metzger2013HeatDiffusion}.}
    \label{tab::sus_props}
\end{table} 

In the heat-pulse experiment  \citep{Metzger2013HeatDiffusion}, the entire system was initially at a uniform temperature of $T_\mathrm{initial}=293~\si{\kelvin}$. Then, the inner cylinder was heated to $T_\mathrm{in}=298~\si{\kelvin}$ for a duration of $5~\si{\second}$, while the outer cylinder was kept at $T_\mathrm{out}=293~\si{\kelvin}$. After that, the heater was turned off, and the temperature of inner cylinder decayed back to $293~\si{\kelvin}$. This heating and cooling process was performed for both the unsheared and sheared (varying inner cylinder rotation rate) suspensions. For our calibration, we use the case of $\dot{\gamma} = 10~\si{\per \second}$. Next, we describe how the heat-pulse experiment was simulated using the TFM to calibrate our proposed model. 

\begin{figure}
    \centering
    \includegraphics[width=0.9\linewidth]{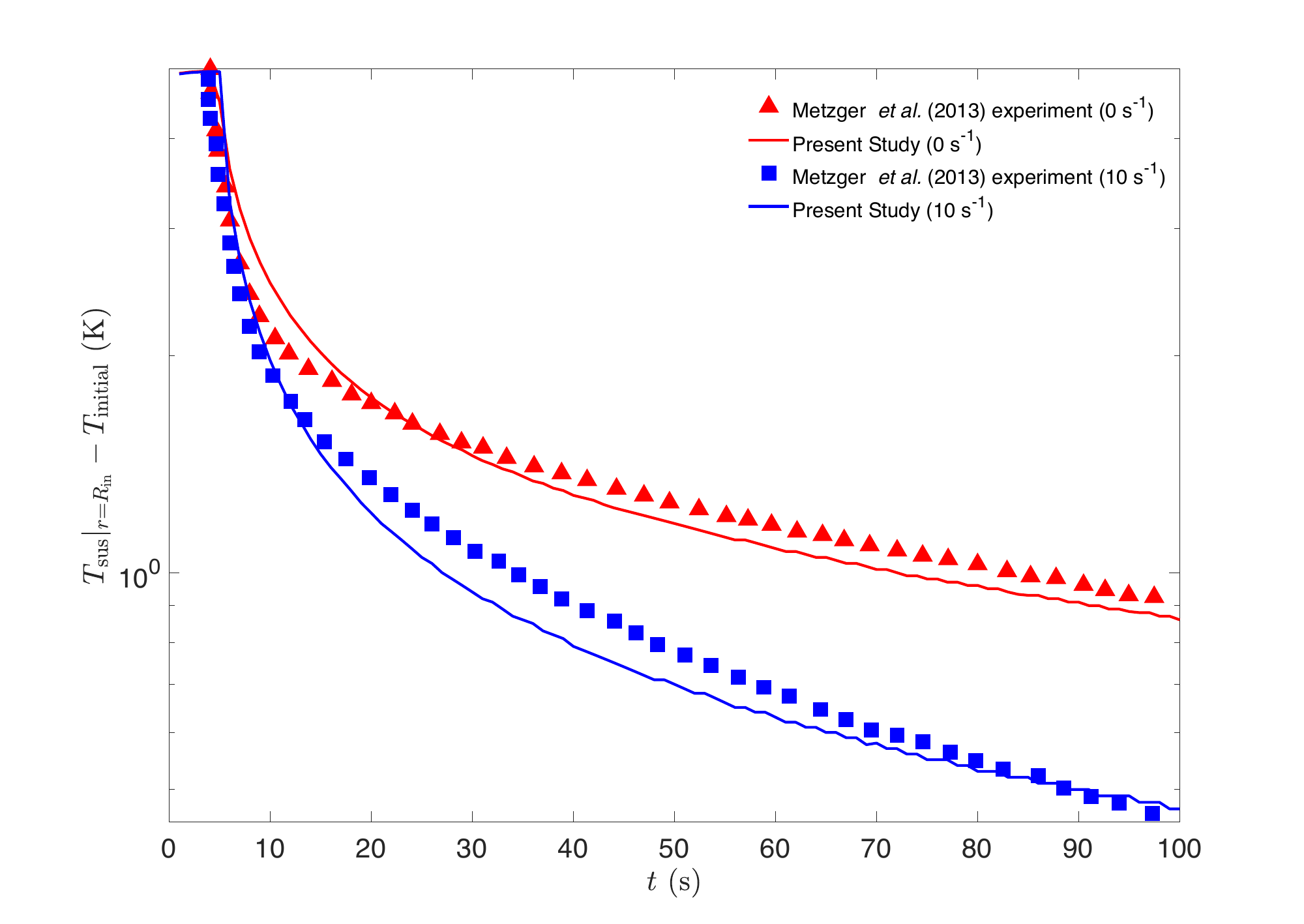}
    \caption{Unsteady temperature response  $T_\mathrm{ sus}|_{r=R_\mathrm{in}}$ at the inner cylinder ($r=R_\mathrm{in}$) for the heat-pulse experiment, starting from a uniform temperature $T_{\rm initial} = 293~\si{\kelvin}$. This plot shows the successful calibration the TFM to the experimental data of \citet{Metzger2013HeatDiffusion}. The calibration against the  $\dot{\gamma}=10~\si{\per\second}$ experimental data yields $\beta = 0.2$ and $m = 1$ (see Appendix~\ref{cal:shear_app}). The case of $\dot{\gamma}=0~\si{\per\second}$ is provided for completeness only and discussed in Appendix~\ref{cal:no_shear}.}
    \label{fig::Val_CC}
\end{figure}

The BCs are specified in Table~\ref{tab::BC}. The outer cylinder was approximated as insulated and always held stationary. During the 5~\si{\second} of heating, the inner cylinder was held stationary and the wall temperature was fixed at $298$~\si{\kelvin} in our model; again, this neglects the heating time for the cylinder to reach the set temperature, but agrees with previous modeling work \citep{Dbouk2018HeatSimulation}. During the cooling process, we approximated the BC at the inner cylinder as adiabatic, which neglects the energy storage term in the inner cylinder as it cools. Because the heat transfer depends on shearing and the parameters may vary with time, which cannot be established for the published experimental description, we instead apply an adiabatic BC during cooling for the inner cylinder (similar to the approach in  \cite{Dbouk2018HeatSimulation}). Thus, the only free parameters in this problem are $\beta$ and $m$ in the shear-dependent inter-phase heat transfer coefficient $K_h$ from Eq.~\eqref{eq::interPhaseHeatTransferCoeff}.

By comparing our model's prediction to the experimental data from \citep{Metzger2013HeatDiffusion}, as shown in Fig.~\ref{fig::Val_CC}, we calibrated the parameters, yielding $\beta = 0.2$ and $m = 1$ (see Appendix~\ref{cal:shear_app}). Note that these parameters are not the same as found by \citet{Metzger2013HeatDiffusion}, who determined $\beta = 0.046$ in their expression for the shear-dependent effective diffusivity. Of course, since the TFM captures different physics (recall the discussion in Section~\ref{sec:Math_HT}), it is not expected that the fitting parameter values would be the same. Note that $m = 1$ indicates that inter-phase heat transfer coefficient $K_h$ depends linearly on the particle-based thermal P\'eclet number $Pe_\mathrm{th}$ (recall Eq.~\eqref{eq::interPhaseHeatTransferCoeff}), which is generally expected \cite{Zydney1988AugmentedSuspension}. 

\begin{table}
    \centering
    \begin{tabular}{l||l|l|l}
    Location & BC & $t<5$~\si{s} (heating) & $t>5$~\si{s} (cooling)\\
    \hline
    \hline
    \multirow{2}{*}{Inner cylinder} & Thermal & $T_\mathrm{in}=298$ \si{\kelvin} & Adiabatic\\
    & Shear rate & $\dot\gamma=0$ \si{s^{-1}} & $\dot\gamma=10$ \si{s^{-1}}\\
    \hline
    \multirow{2}{*}{Outer cylinder} & Thermal & Adiabatic & Adiabatic \\
    & Shear rate & $\dot\gamma=0$ \si{s^{-1}} & $\dot\gamma=0$ \si{s^{-1}}\\
    \end{tabular}
    \caption{Boundary conditions for the TFM simulations of the Couette cell heat-pulse experiment. The adiabatic condition enforces $\bm{\nabla} T_f \cdot\bm{n}= \bm{\nabla} T_p \cdot\bm{n}= 0$ on the appropriate boundaries with unit normal $\bm{n}$. The specified shear rate is estimated as  $\dot{\gamma}= (2 \Omega_\mathrm{in} R_\mathrm{out}^2) /(R_\mathrm{out}^2 - R_\mathrm{in}^2)$ and achieved by setting a suitable rotation rate of the inner wall $\Omega_\mathrm{in}$.}
    \label{tab::BC}
\end{table}

\subsection{Flow of a sheared suspension in an eccentric Couette cell}
\label{sec:Val_EC}

\begin{figure}
    \centering
    \includegraphics[width=0.8\linewidth]{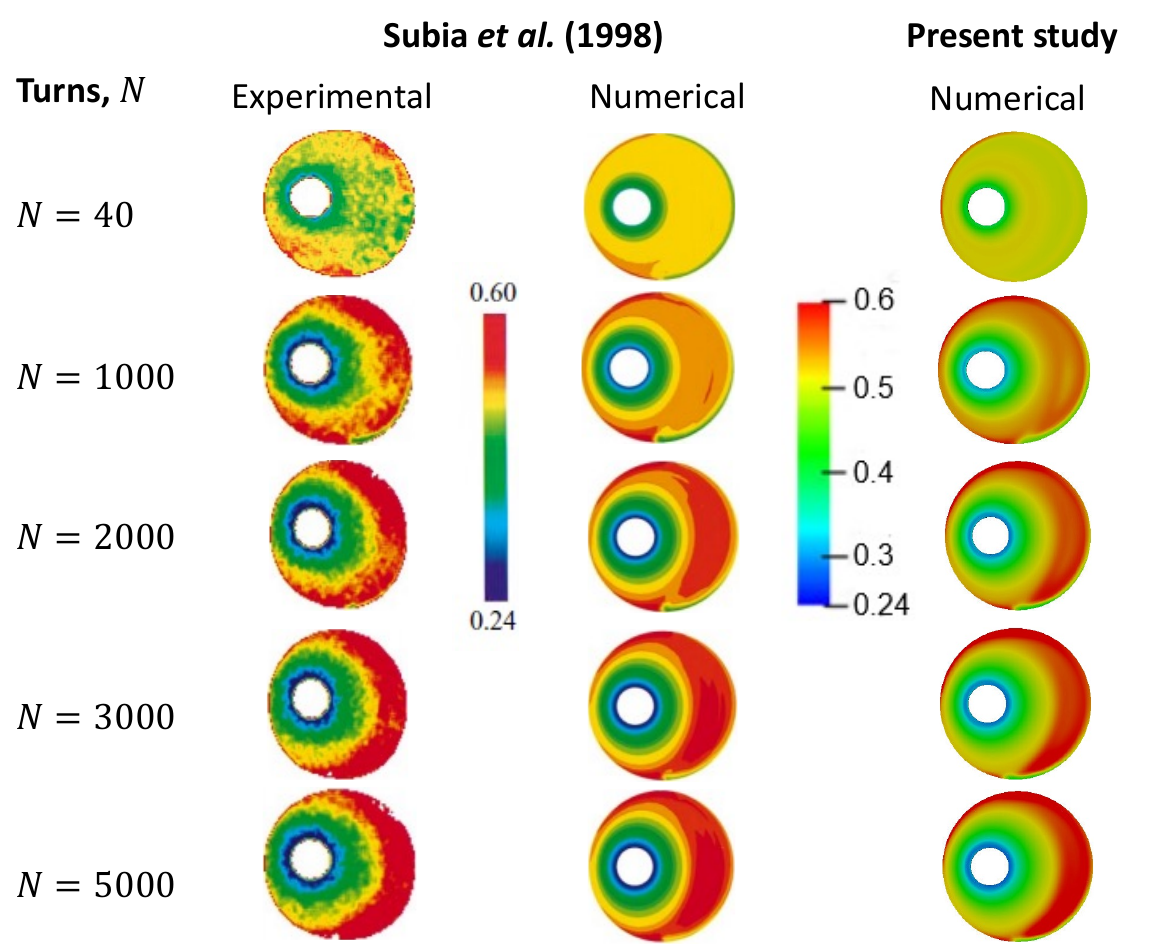}
    \caption{Qualitative validation of the TFM simulation against the results of \citet{Subia1998ModellingEquation} for dense suspension flow (see Table~\ref{tab::subiaParameters} for the parameters) in an eccentric Couette cell.  Color shows contours of the particle volume fraction $\phi$ for different number of turns $N$ of the inner cylinder. The first and second columns are reproduced, with permission, from [Subia, S.R., Ingber, M.S., Mondy, L.A., Altobelli, S.A., Graham, A.L., 1998.\ Modelling of concentrated suspensions using a continuum constitutive equation. Journal of Fluid Mechanics 373, 193–219. doi:10.1017/S0022112098002651 \copyright\ Cambridge University Press.]}
    \label{fig::Val_ECC}
\end{figure}

An eccentric Couette cell is another typical geometry in which flows of dense suspensions are studied. In this geometry, the center of the inner cylinder is offset from the center of the outer cylinder. The eccentricity is quantified by the ratio $E = d/(R_\mathrm{out}-R_\mathrm{in})$, where $d$ is the distance between the centers of the two cylinders. In general, eccentricity can lead to a recirculating region \cite{Ballal1976FlowEffects} that induces additional mixing \cite{Swanson1990AFluids}. \citet{Subia1998ModellingEquation} performed experiments and finite element modeling of suspension flow using the parameters listed in Table~\ref{tab::subiaParameters}. Next, we qualitatively validate our TFM against the numerical model and experiments of \citet{Subia1998ModellingEquation} by plotting $\phi$ contours in Fig.~\ref{fig::Val_ECC} after different number of revolutions (turns) of the inner cylinder. Our TFM results are in good agreement with the previous ones. Note that this validation is for the isothermal case. In  Section~\ref{sec::EccentricDisc}, we explore heat transfer in the eccentric Couette cell.

\begin{table}
    \centering
    \begin{tabular}{l||l|l|l|l|l|l|l|l}
    Quantity & $R_\mathrm{in}$ & $R_\mathrm{out}$ & $E$ & $\Omega_\mathrm{in}$ & $d_p$ & $\mu_f$ & $\phi_b$ & $\rho_f$\\
    \hline
    \hline
    Value & $0.64~\si{\centi\meter}$ & $2.54~\si{\centi\meter}$ & 0.5 & $90~\mathrm{rpm}$ &
    $675~\si{\micro\meter}$ & $4.95~\si{\pascal\second}$ & $0.5$  & $1180~\si{\kilo\gram\per\meter\tothe{3}}$
    \end{tabular}
    \caption{Geometric parameters, particle and fluid properties used in the simulations that validate our TFM against the results of  \citet{Subia1998ModellingEquation}. The suspension is comprised of PMMA particles suspended in a Newtonian fluid.}
    \label{tab::subiaParameters}
\end{table} 

\section{Particle migration in the presence of coupled thermal and shear gradients}
\label{sec:RAndD}

In this section, we discuss our main computational results on the interaction of shear-induced migration with thermal gradients across the system. Specifically, we will consider the two cases in which the temperature difference $\Delta T = T_\mathrm{in}-T_\mathrm{out}$ across the gap is either positive or negative (\textit{i.e.}, heat transfer occurs from the inner to the out wall, or \textit{vice versa}).

\subsection{Concentric Couette cell}
\label{sec:CC}
After calibrating the inter-phase heat transfer coefficient as described in Section~\ref{sec:Val_CC}, we performed a parametric study to understand the interplay between shear and thermal gradients on particle migration. Specifically, we varied the bulk particle volume fraction $\phi_b$, the thermal P\'eclet number $Pe_\mathrm{th} = \dot{\gamma} d_p^2 / \alpha_p$ and the temperature difference $\Delta T$ across gap. As in the calibration described in Section~\ref{sec:Val_CC}, we use the concentric Couette cell geometry shown in Fig.~\ref{fig::geoMesh}. 

Keeping in mind the salient application of suspension flows to electronics cooling \citep{Dbouk2019AParticles}, in these simulations we consider boron nitride (BN) particles dispersed in a fluoro-carbon (FC) fluid. This choice of suspension is to ensure that the particles and fluid have nearly identical densities, and hence, the suspension is neutrally buoyant. The particles have constant thermophysical properties as given in Table~\ref{tab::sus_props_BN_FC}. 
The thermophysical properties of the FC-43 fluid are temperature dependent \cite{ElectronicsMaterialsSolutionsDivision20203MFC-43}. We fit a fourth-order polynomial to the data for the FC-43 fluid viscosity as a function of temperature from \cite{ElectronicsMaterialsSolutionsDivision20203MFC-43}, as given in Table~\ref{tab::sus_props_BN_FC}. 
The choice of such particles and fluid is due to large contrast in their thermal conductivities, which we hypothesize will allow us to observe significant thermal performance enhancement. The thermophysical properties of the particles are taken to be constant because measurements of the temperature-dependence of $C_{p,p}$, $k_p$, and $\rho_p$ for BN show that it is quite weak over the temperature ranges ($\approx 200$--$400~\si{\kelvin}$) explored in this work \cite{Shipilo1986ElectricalBeta-BN,Solozhenko1997CompressionK,Gavrichev1993Low-temperatureModifications}.

\begin{table}
    \centering
    \begin{tabular}{l||l|l}
      Property  & Particle (`$p$') & Fluid (`$f$')\\
      \hline
      \hline
      Density $\rho$ (\si{\kilo\gram\per\meter\tothe{3}}) & 1900 & $2508 - 2.18T_f$ \\
      \hline
      Dynamic viscosity $\mu$ (\si{\pascal\second}) & -- & \multirow{3}{*}{\shortstack{$0.3933 - 0.0035T_f + 
      (1.13\times 10^{-5})T_f^2$\\$- (1.5\times 10^{-8})T_f^3 + (6.7\times 10^{-12})T_f^4$}} \\
      &\\
      &\\
      \hline
      Thermal conductivity $k$ (\si{\W\per\m\per\kelvin}) & 35.5 & $0.08611 - (7\times 10^{-5})T_f$ \\
      \hline
      Specific heat $C_p$ (\si{\J\per\kg\per\kelvin}) & 960 & $589.8 + 1.554T_f$ \\
    \end{tabular}
    \caption{Thermophysical properties of the suspension of boron nitride (BN) particles into a fluoro-carbon (FC) fluid \cite{ElectronicsMaterialsSolutionsDivision20203MFC-43}. For the temperature-dependent fluid properties' formulas, the coefficients are in the appropriate (implied) SI units.}
    \label{tab::sus_props_BN_FC}
\end{table} 

\begin{table}
    \centering
    \begin{tabular}{l||c|c|c}
      Temperature BC  & Notation & $T_\mathrm{in}$ & $T_\mathrm{out}$\\
      \hline
      \hline
      Case 1: migration with thermal gradient & $\Delta T > 0$ & 323~\si{\kelvin}  & 293~\si{\kelvin} \\
      \hline
      Case 2: migration against thermal gradient & $\Delta T < 0$ & 293~\si{\kelvin} & 323~\si{\kelvin} \\
    \end{tabular}
    \caption{Temperature boundary conditions for the inner (`in') and outer (`out') cylinders of the Couette cell.}
    \label{tab::BC_CC}
\end{table} 

In these simulations, we use BN particles with $d_p = 675~\si{\micro\meter}$ at an initial (spatially uniform) volume fraction of $\phi_b=0.5$, unless otherwise stated. The inner cylinder is rotated so as to maintain $\dot\gamma = 3~\si{\per\second}$ at steady state. The outer cylinder is held stationary. Consequently, due to the shearing, particles are expected to migrate from the inner cylinder towards the outer one. We consider two different sets of thermal BCs as described in Table~\ref{tab::BC_CC}. In doing so, we wish to characterize the radial particle migration fields when the shear-induced particle migration and heat transfer are in same ($\Delta T > 0$) vs.\ opposite ($\Delta T < 0$) directions across the gap.

Having specified the flow conditions, we can now justify why viscous dissipation is neglected in the energy equations (Eqs.~\eqref{eq::energy_p} and \eqref{eq::energy_f}). Specifically, the Eckert number for this flow is $Ec = U_c^2/(C_p \Delta T) \simeq 10^{-11}$, indicating that conduction is the dominant heat transfer mechanism in this system. Here, we have taken the characteristic velocity $U_c$ to be the inner wall velocity (corresponding to the $3~\si{\per \second}$ shear rate), and $C_p=\phi_b C_{p,p} + (1-\phi_b)C_{p,f}$ is the suspension's specific heat estimated as the bulk-volume-fraction weighted average at the initial temperature of $293~\si{\kelvin}$.

\subsubsection{Interplay between heat transfer and shear migration}
\label{sec:interplay_results}

\begin{figure}
\centering
\includegraphics[width=0.5\linewidth]{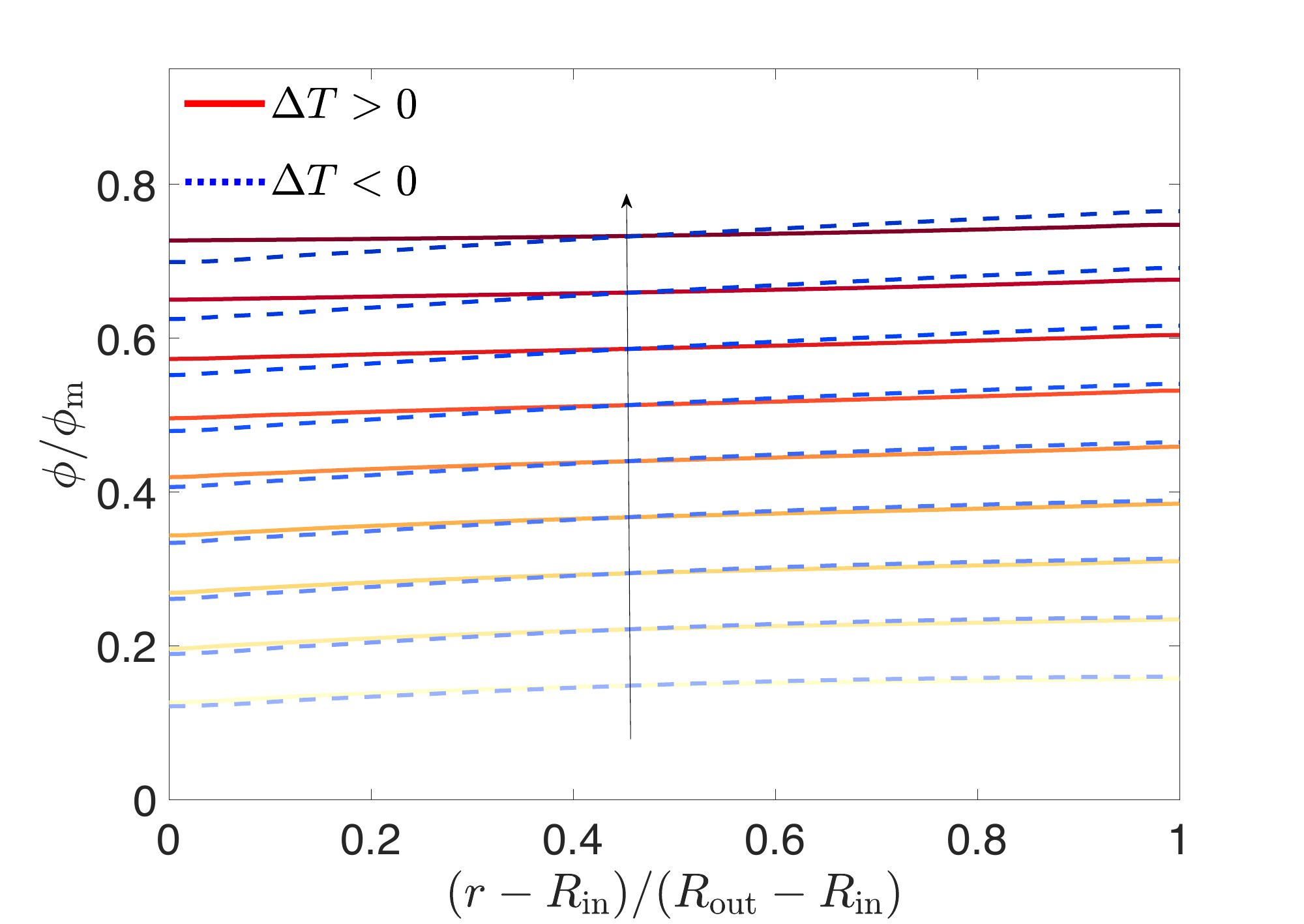}
\caption{Effect of bulk particle volume fraction $\phi_b$ on the radial distribution of particles $\phi$ for the concentric Couette cell for $d_p = 675~\si{\micro\meter}$ and $ \dot {\gamma} = 3~\si{\per\second}$. Arrows indicate the direction of increasing $\phi_b$.}
\label{fig::bulkVolumeFraction}
\end{figure}

The impact of the direction of the temperature gradient across the Couette cell on the radial particle migration profiles at different bulk volume fractions ($\phi_b = 0.1$ to $\phi_b = 0.5$) is shown in Fig.~\ref{fig::bulkVolumeFraction}. For $\Delta T < 0$, the thermal and shear gradients lead to migration fluxes in the same direction and, hence, enhance migration (in comparison to $\Delta T > 0$, for which the fluxes are in opposite directions). Therefore, at each $\phi_b$, we observe particle migration towards the outer wall (segregation of the mixture). From Fig.~\ref{fig::bulkVolumeFraction}, we also observe that the particle distribution profiles are similar for both $\Delta T>0$ and $\Delta T<0$ for small $\phi_b$, while a prominent difference emerges as $\phi_b$ increases. Therefore, the interplay between shear and thermal gradients is more pronounced for dense suspensions.

\begin{figure}
\centering
\begin{subfigure}{.5\linewidth}
    \includegraphics[width=\linewidth]{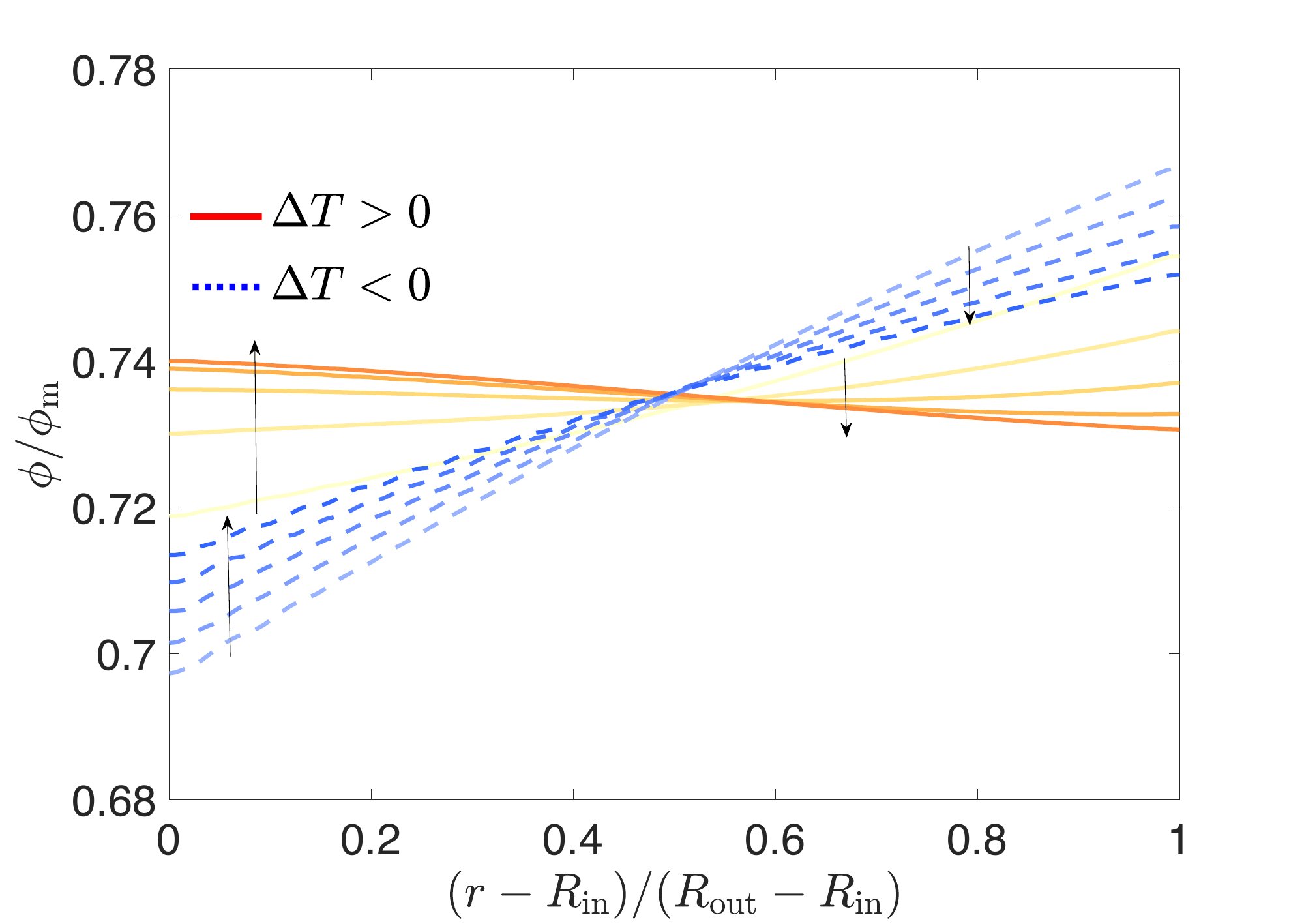}
    \caption{}
    \label{fig:phivsr_dp}
\end{subfigure}\hfill
\begin{subfigure}{.5\linewidth}
    \includegraphics[width=\linewidth]{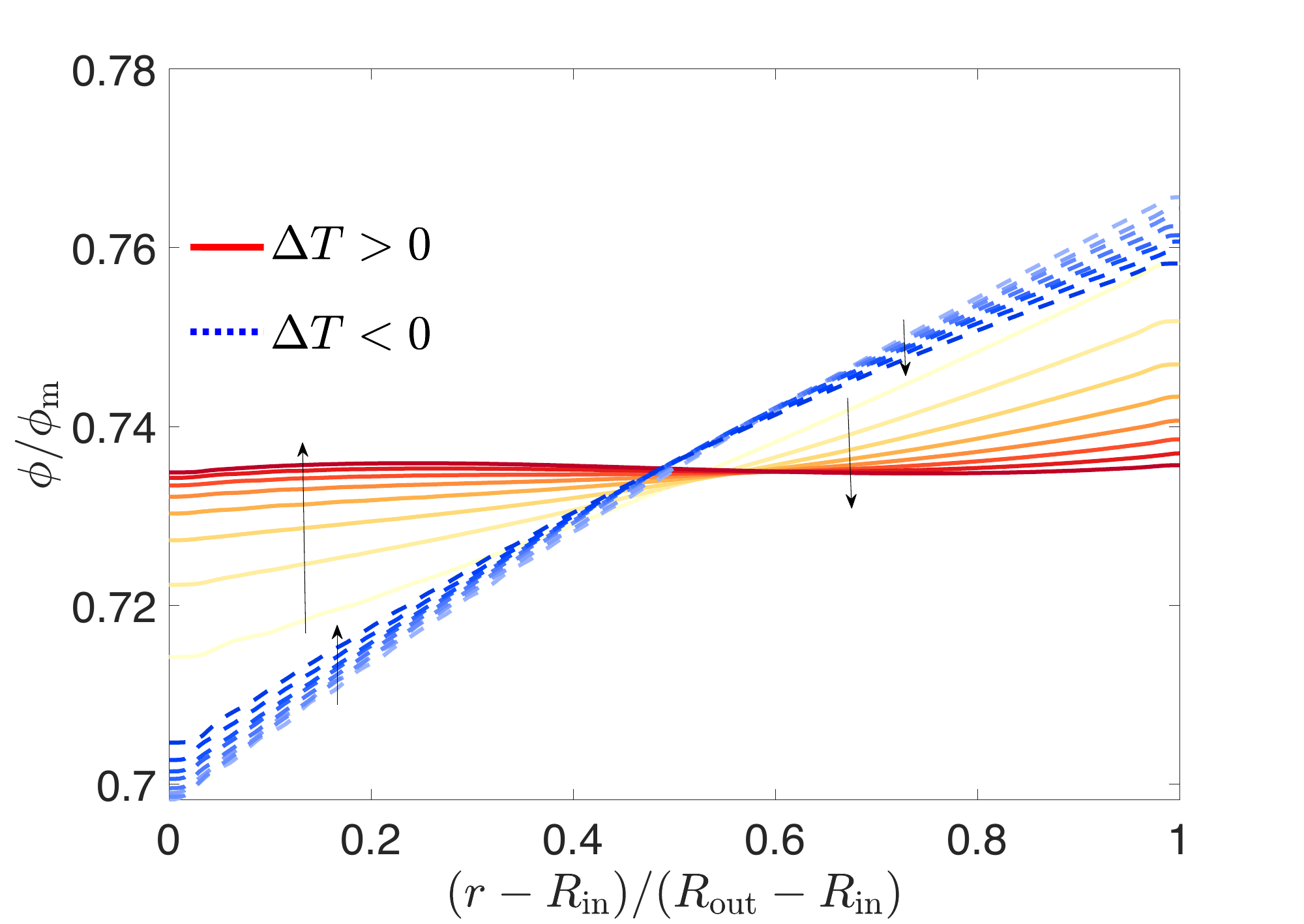}
    \caption{}
    \label{fig:phivsr_gamma}
\end{subfigure}
\caption{Effect of varying the thermal P\'eclet number $Pe_\mathrm{th}=\dot{\gamma} d_p^2 / \alpha_p$ via (a) the particle diameter $d_p$ and (b) the imposed shear rate $\dot\gamma$ (by rotation of the inner cylinder) on the radial distribution of particles $\phi$ in the concentric Couette cell for $\phi_b=0.5$. Arrows indicate the direction of increasing $d_p$ or $\dot\gamma$.}
\label{fig::thermalPe}
\end{figure}

To fully explore the possible behaviors, we next analyze the impact of thermal P\'eclet number $Pe_\mathrm{th}$ on the particle migration. We vary $Pe_\mathrm{th}$ by changing either $\dot\gamma$ or $d_p$: we performed simulations with $d_p = 0.5-1.5~\si{mm}$ (Fig.~\ref{fig:phivsr_dp}) $\dot \gamma = 1-8~\si{\per \second}$ (Fig.~\ref{fig:phivsr_gamma}), for both $\Delta T > 0$ and $\Delta T < 0$.  For both the temperature BCs, particle migration is suppressed at larger $d_p$ and $\dot\gamma$ ($\Rightarrow$ larger $Pe_\mathrm{th}$). Moreover, $\Delta T > 0$ results in a more homogeneous suspension because the contribution of the thermal gradients to the migration flux opposes that of the shear gradients. Significant particle segregation across the gap is observed for $\Delta T < 0$ because the fluxes due to thermal and shear gradients enhance each other, which aids particle migration. When $\Delta T > 0$, almost no particle migration is observed for the strongly sheared suspensions with largest $Pe_\mathrm{th}$ (corresponding to, \textit{e.g.}, $d_p = 0.5~\si{mm}$ and $\dot \gamma = 8~\si{\per \second}$).

\subsubsection{How the thermo-rheological fluxes affect particle migration}

\begin{figure}
    \centering
    \includegraphics[width=0.5\linewidth]{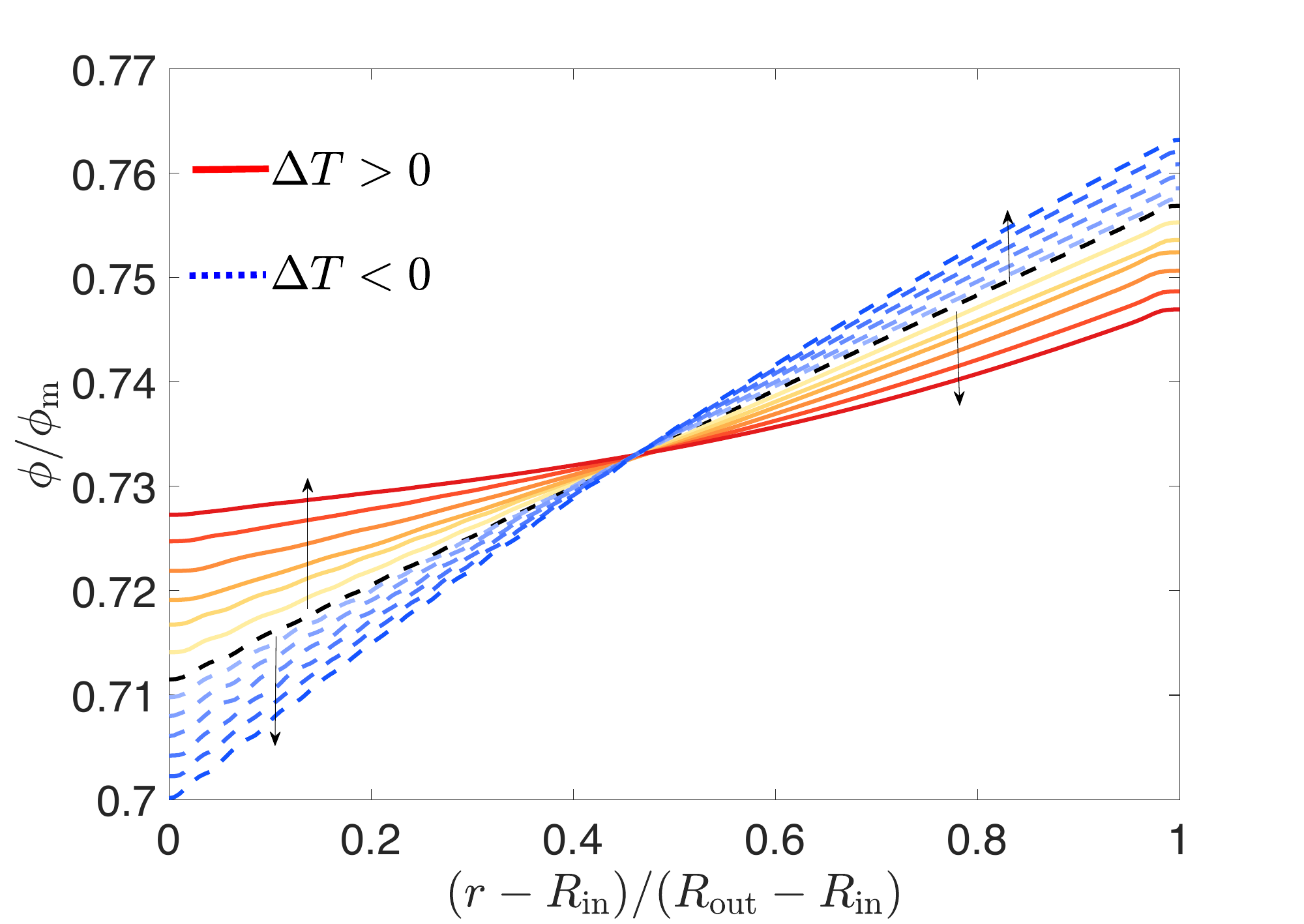}
    \caption{Effect of the temperature difference $\Delta T$ across the gap on the radial distribution of particles $\phi$ in the concentric Couette cell for $\phi_b=0.5$, $ \dot {\gamma} = 3~ \si{\per\second}$ and $d_p = 675~\si{\micro\meter}$. Arrows indicate the direction of increasing $|\Delta T|$. The black dashed curve represents the isothermal case ($\Delta T=0$).}
    \label{fig::DeltaT}
\end{figure}

In Section~\ref{sec:thermo-rheol}, we provided an abstract discussion on the origin of the thermo-rheological particle migration fluxes within the TFM framework. In this subsection, we quantify the effect of temperature gradients on the particle migration profiles for a fixed shear gradient. We now subject the Couette cell  to $|\Delta T| = 5 - 30~\si{\kelvin}$ for each temperature BC. Figure~\ref{fig::DeltaT} shows that, for the BC with $\Delta T > 0$, as the temperature difference increases, particle migration is reduced due to the opposing shear- and thermal-driven particle migration fluxes. The black dashed curve in Fig.~\ref{fig::DeltaT} represents the isothermal case in which the migration is solely due to the shear gradient. For BC with opposing heat transfer ($\Delta T < 0$), the contribution of the thermal-gradient-induced particle flux aids the shear-gradient-induced one, leading to enhanced particle migration towards the outer wall. As $|\Delta T|$ is increased (in this $\Delta T < 0$ case), the augmentation from the thermal gradients on the overall migration flux is evidently stronger as well.

\subsection{Eccentric Couette cell}
\label{sec::EccentricDisc}

For the concentric Couette cell, the heat transfer characteristics are dominated by the suspension's effective thermal conductivity, showing only weak dependence on the various parameters varied, which is why we did not discuss this point in Section~\ref{sec:CC}.  A more interesting setup in which to characterize the overall system's thermal performance is the dense suspension flow in an eccentric Couette cell (recall Section~\ref{sec:Val_EC}). For example, eccentric Couette cells are an effective way to enhance mixing at low Reynolds number \cite{Swanson1990AFluids} due to the added complexity of the recirculating regions.

We quantify the system's heat transfer by calculating the heat transfer coefficient $h_\mathrm{in} = q''/(T_\mathrm{in}-T_\mathrm{out})$, where $q''$ is the heat flux from the inner cylinder. Note that this convection coefficient $h_\mathrm{in}$, characterizing the heat transfer in the entire system, is not the same as inter-phase heat transfer coefficient $K_h$ from Eq.~\eqref{eq::interPhaseHeatTransferCoeff}, which captures heat transfer between the particles and fluid phases. Then, making $h$ dimensionless, we calculate the Nusselt number $Nu$ of the system. In this section, we show the dependence of $Nu$ on the eccentricity $E = d/(R_\mathrm{out}-R_\mathrm{in})$ of the Couette cell. 

We use the basic Couette cell geometry from Section~\ref{sec:Val_EC} with varying eccentricities from $E=0$ to $0.6$ (above which we observe jamming of particles). The eccentricity in the geometry renders the mesh non-orthogonal. Hence, additional non-orthogonality correction loops are run in each time step in order to obtain accurate results \cite{MNC19}.  We focus on the neutrally buoyant suspension of BN particles with $d_p=675$~\si{\micro\meter} and $\phi_b=0.3$ in an FC fluid. The thermophysical properties of the two phases are given in Table \ref{tab::sus_props_BN_FC}, as before. Finally, the same sets of thermal BCs as given in Table~\ref{tab::BC_CC} are simulated to highlight the differences between the case when the thermal gradient across the gap aides or opposes the shear-induced particle migration.

As mentioned above, our figure of merit for quantifying heat transfer is the system's Nusselt number $Nu$. It is calculated by applying an energy balance at the surface of the hotter cylinder. For the case 1 BC, at steady state, this energy balance on the inner cylinder wall gives
\begin{equation}
    q''|_{r=R_\mathrm{in}} = - \left[k_p \phi  \frac{dT_p}{dr}  + k_f (1-\phi) \frac{dT_f}{dr}\right]_{r=R_\mathrm{in}} = h_\mathrm{in}(T_\mathrm{in} - T_\mathrm{out}).
    \label{eq::system_htc}
\end{equation}
The second equality above is used to calculate the heat transfer coefficient $h_\mathrm{in}=h|_{r=R_\mathrm{in}}$ for the case 1 BC. Similarly, $h_\mathrm{out}=h|_{r=R_\mathrm{out}}$ for case 2 BC is calculated by applying the same energy balance now at the stationary outer cylinder ($r=R_\mathrm{out}$). For comparison purposes, $h_\mathrm{in}$ is also calculated for the case 2 BC from $h_\mathrm{out}$ via the steady energy balance as $h_\mathrm{in} R_\mathrm{in} = h_\mathrm{out} R_\mathrm{out}$. In Eq.~\eqref{eq::system_htc}, the particle volume fraction and temperature gradients at the hotter wall are obtained from the TFM simulation. Finally, $Nu = h_\mathrm{in} (R_\mathrm{out} - R_\mathrm{in}) / k_\mathrm{sus}$ (defined the same way for all eccentricities), where the suspension thermal conductivity is taken as $k_\mathrm{sus} = \phi_b k_p + (1 - \phi_b) k_f$ for the purposes of computing $Nu$. The Nusselt number is a `coarse' measure of the system heat transfer, thus in our definition of it, we do not account for the shear-induced migration explicitly. Furthermore, since $k_p \approx 500 k_f \gg k_f$, $q''$ from Eq.~\eqref{eq::system_htc} is dominated by the particle heat flux. Hence, the particle phase's heat fluxes at the inner (for $\Delta T > 0$) and outer (for $\Delta T < 0$) cylinders drive the thermal performance of the system.

\begin{figure}
\centering
\begin{subfigure}{.5\textwidth}
  \centering
  \includegraphics[width=\linewidth]{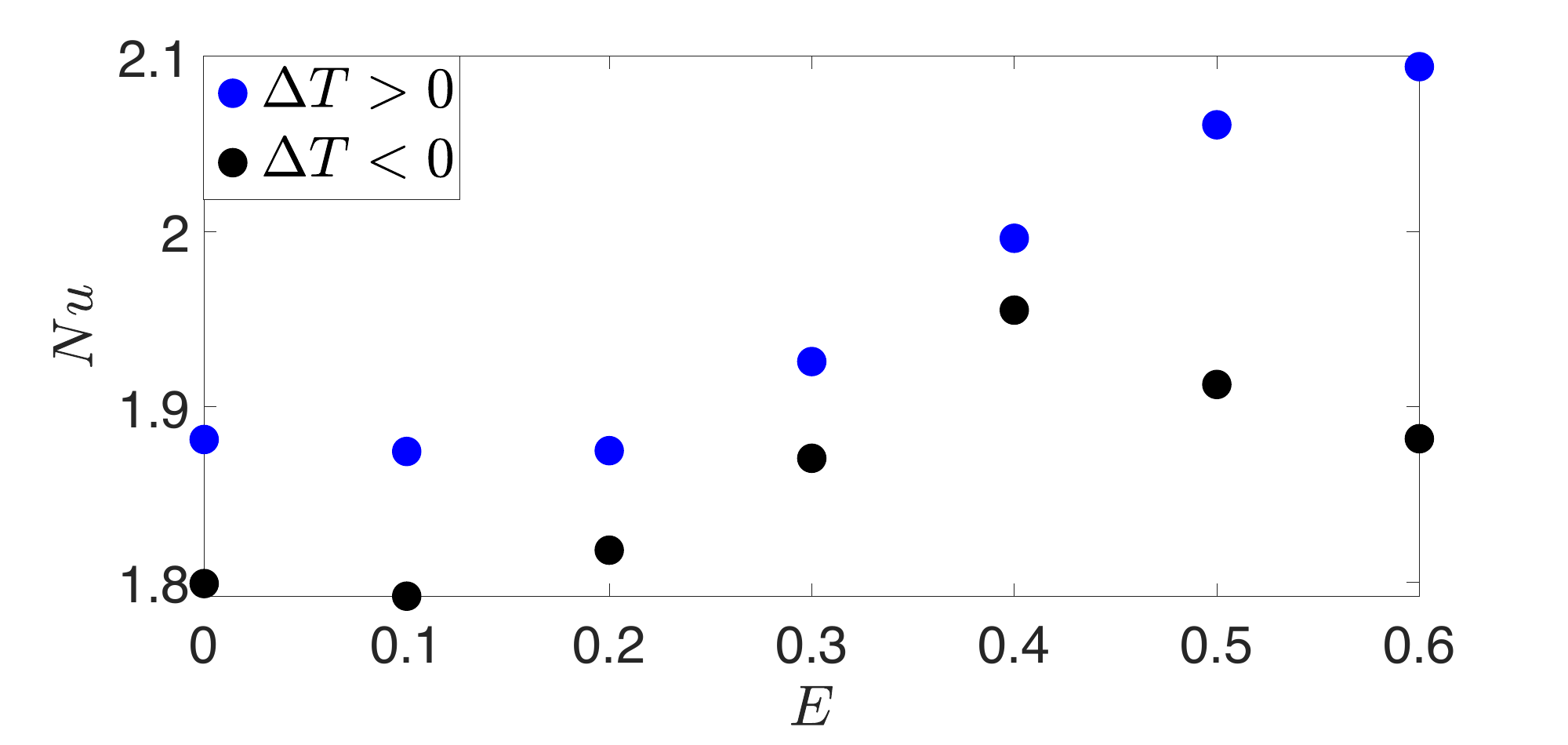}
  \caption{}
  \label{fig:NuvsE}
\end{subfigure}\hfill
\begin{subfigure}{.5\textwidth}
  \centering
  \includegraphics[width=\linewidth]{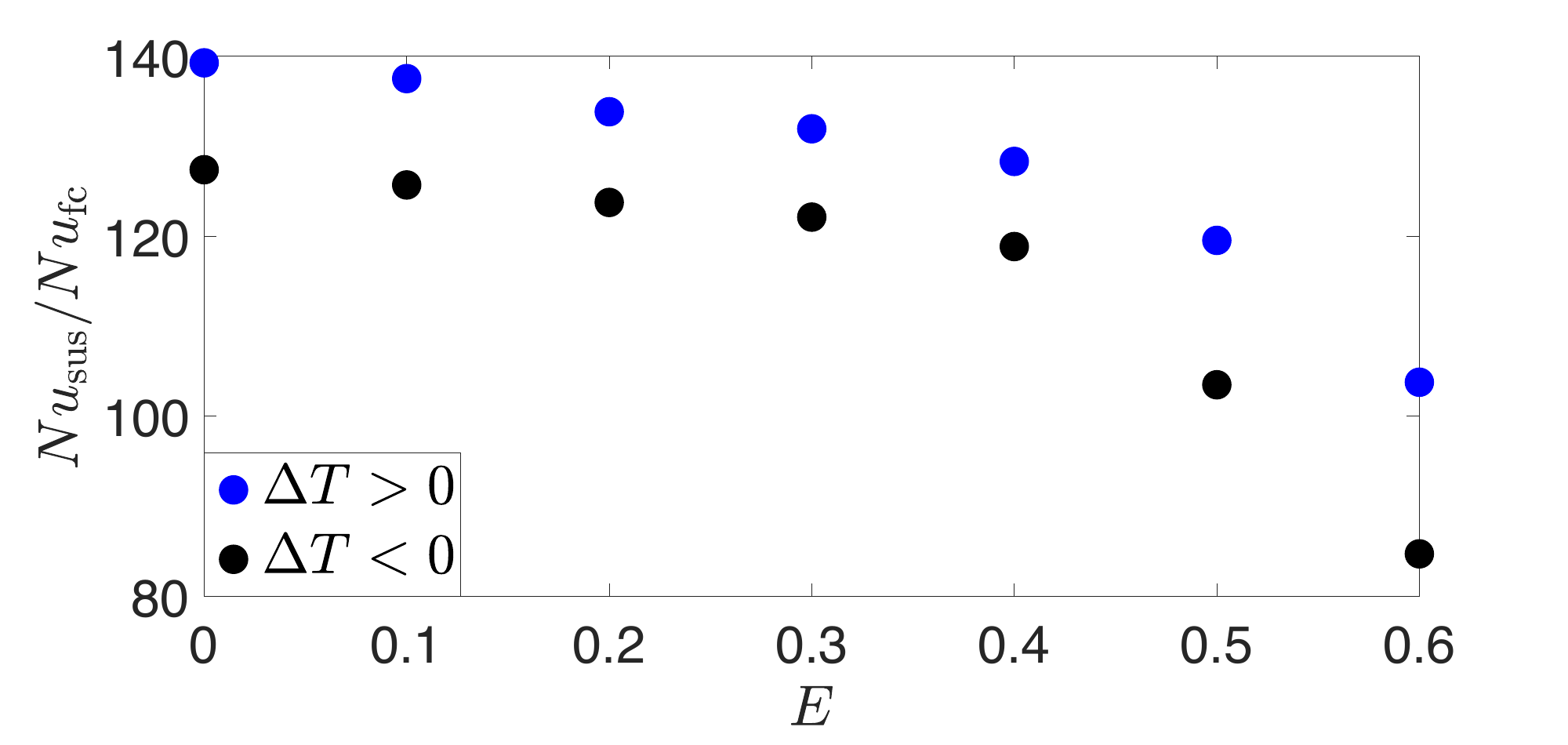}
  \caption{}
  \label{fig:NuratiovsE}
\end{subfigure}
\caption{Effect of Couette cell eccentricity $E$ on (a) Nusselt number $Nu$ of the suspension and (b) the Nusselt enhancement factor $Nu_\mathrm{sus}/Nu_\mathrm{fc}$ (with respect to clear FC-43 fluid with no particles) for $\phi_b = 0.3$, $\dot{\gamma} = 3~\si{\per\second}$ and $d_p = 675~\si{\micro\meter}$.}
\label{fig::eccentricity}
\end{figure}

\begin{figure}
    \centering
    \includegraphics[width=\linewidth]{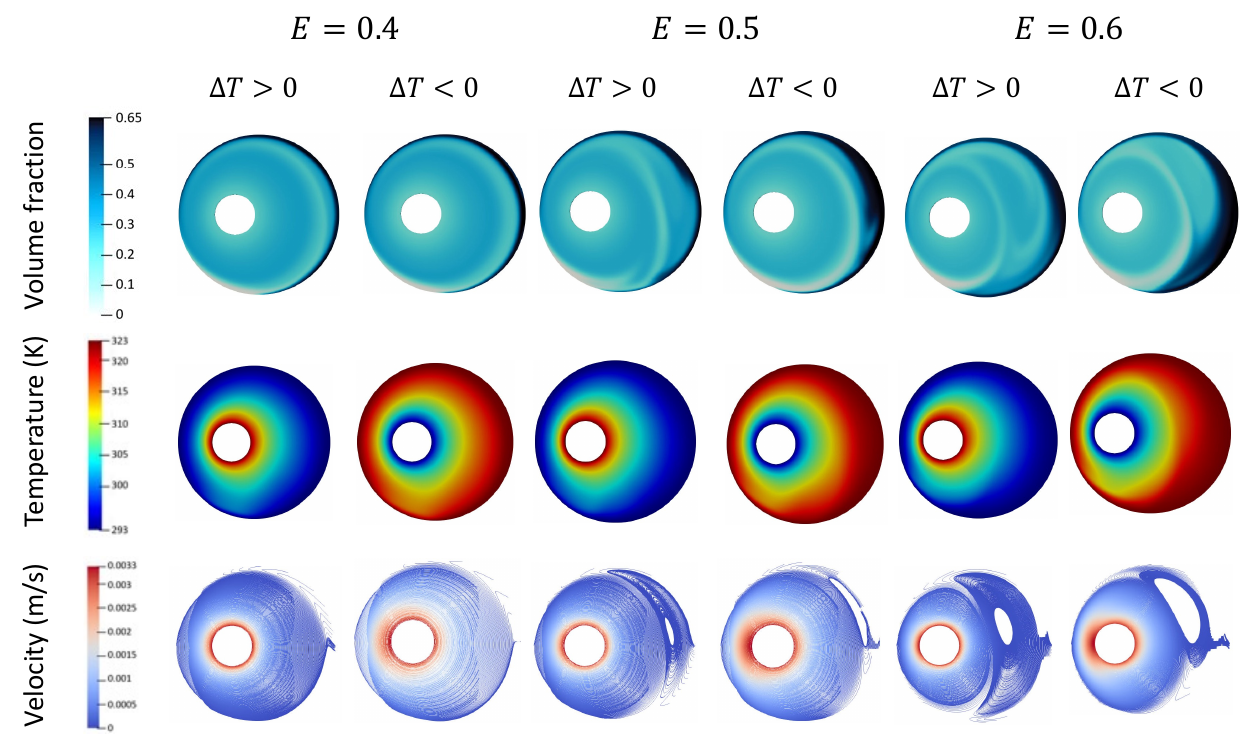}
    \caption{Comparison of the volume fraction, suspension temperature field, and suspension velocity (visualized by streamlines color-coded by the velocity magnitude) for different $E$ in the eccentric Couette cell flow, considering both types of thermal BCs.}
    \label{fig::E_Table}
\end{figure}

Figure~\ref{fig:NuvsE} shows that the Nusselt number increases with the eccentricity of the cell for $\Delta T > 0$ case, while an optimum at $E=0.4$ is observed for $\Delta T < 0$ case. 
The Nusselt number for both thermal BCs is comparable up to $E=0.4$, after which $\Delta T > 0$ performs better than $\Delta T < 0$ for $E=0.5$ and $E=0.6$. This behavior can be explained by considering the results shown in Fig.~\ref{fig::E_Table}. Differences in the particle migration fields are observed for $E=0.5$ and $E=0.6$ between the two thermal BCs (especially near the outer wall in the wider gap), while the particle migration fields for $E=0.4$ are visually indistinguishable for both BCs. 

For $E=0.5$ and $E=0.6$ with $\Delta T<0$, particle migration is more pronounced than for $\Delta T>0$, as evidenced by the larger variation in the volume fraction along the outer wall. The additional migration for $\Delta T < 0$ is due to shear and thermal gradients aiding each other (to increase the overall particle flux). This migration, in turn, decreases the particle flux term in Eq.~\eqref{eq::system_htc} (for $E=0.5$ and $E=0.6$ with $\Delta T<0$) and, hence, the Nusselt number is reduced compared to the $\Delta T > 0$ case. For this same case, particle migration is also enhanced by the recirculating flow (a `vortex') observed in the velocity fields in Fig.~\ref{fig::E_Table}. For eccentricity ratio $E\lesssim 0.4$, the recirculating flow is not observed (consistent with the known theory for a Newtonian eccentric Couette flow \cite[Fig.~17]{Ballal1976FlowEffects}), therefore particle migration is diminished for $E=0.4$ compared to $E=0.5$ and $E=0.6$. In addition, the vortex becomes larger and moves towards the outer wall for $\Delta T<0$ (compared to $\Delta T>0$). Once again, this observation highlights the interplay between heat transfer and particle migration, further suggesting that the flow characteristics can also be tuned in this system via said interplay.

Figure~\ref{fig:NuratiovsE} shows the dependence of the enhancement factor $Nu_\mathrm{sus}/Nu_\mathrm{fc}$ (with respect to clear FC-43 fluid with no particles) on the eccentricity ratio $E$. Even though $Nu_\mathrm{sus}$ (up to $E=0.4$ for the $\Delta T < 0$ case) and $Nu_\mathrm{fc}$ are both augmented at higher $E$, the enhancement factor decreases with $E$. This observation suggests that there is a trade-off in the eccentric Couette cell flow. The suspension yields the largest enhancement (compared to the clear FC fluid) for the concentric case ($E=0$). The relative improvement is reduced thereafter, although it is still significant. The reason for the diminished improvement with increasing $E$ is that the suspension entrains particles near the outer wall in the wider gap section (as evident from the top row of Fig.~\ref{fig::E_Table}), whereas the clear FC fluid flow (having no particles) does not exhibit this pathology. Therefore, $Nu_\mathrm{sus}/Nu_\mathrm{fc}$ shows a mild decreasing trend with $E$, as the accumulation of particles near the outer wall cancels out some of the heat transfer enhancement enabled by the highly-conductive BN particles.

Considerable deviation between the two BCs is observed at $E=0.5$ and $E=0.6$ due to the enhanced particle migration for the case 2 BC resulting in lower $Nu_\mathrm{sus}$ and hence lower $Nu_\mathrm{sus}/Nu_\mathrm{fc}$.

\section{Conclusion}
\label{sec:Conclusion}

In the present work, we developed a two-fluid model (TFM) for simulation of heat transfer in dense non-Brownian suspensions. We used this TFM to investigate the effect of combined shear and thermal gradients on the phenomenon of particle migration in a Couette cell. We built upon our previous work, \citet{MNC19}, in which we focused only on the anisotropy stress tensor (recall Eq.~\eqref{eq::Sigma_s}) and the shear-induced particle migration aspect. In this work, we extended the latter by  calibrating a closure relation for the inter-phase heat transfer coefficient, given in Eq.~\eqref{eq::interPhaseHeatTransferCoeff}, which takes into account the joint effect of the particle thermal diffusivity and rate-of-strain tensor of the particulate phase on the inter-phase (fluid-particle) heat transfer. In this respect, unlike previous models, the proposed TFM allows for thermal disequilibrium between the phases. Specifically, in the TFM, shearing the suspension enhances the inter-phase heat transfer coefficient, rather than the intrinsic thermal conductivity of particles (as in previous models), which should be a fixed thermophysical quantity. This approach allowed us to explain the origin of a novel thermo-rheological flux and its effect on the particle migration phenomenon. Specifically, Eq.~\eqref{eq::stress_force_x_final} shows that a flux due to thermal gradients can act to oppose a flux generated by shear gradients.

To further understand the interplay of shear and thermal gradients on particle migration, we conducted a parametric study of dense suspension flow in a Couette cell by varying $\phi_b$, $Pe_{\rm th}$ (by varying $\dot{\gamma}$ and $d_p$) and $\Delta T$ across the gap. An enhanced particle segregation is seen when $\Delta T < 0$ because the signs of the shear and thermal gradients term in Eq.~\eqref{eq::stress_force_x_final} are the same (fluxes are in the same direction). On the other hand, for $\Delta T > 0$, the fluxes due to shear and thermal gradients oppose each other resulting in a homogeneous particle distribution across the gap. The difference in particle migration profiles between the two cases increase as the suspension is made denser. This observation follows from the fact that the effect of individual migration flux terms is  more pronounced when more particles (by volume) are added to the system. In addition, we observed that particle migration is reduced for both thermal BCs in strongly sheared suspensions (larger thermal P\'eclet number). 

Moreover, for the $\Delta T > 0$ case, as $|\Delta T|$ across the gap is increased, the opposing thermal-gradient-induced flux strengthens, and ultimately cancels out more of the shear-gradient-induced flux. The result is a more homogeneous suspension across the gap. Therefore, from a practical point of view, to reduce particle migration and improve the thermal performance of this flow system, it is recommended that the Couette cell is densely filled (say, $\phi_b = 0.5$) with large particles (say, $d_p = 0.5~\si{mm}$) and subjected to high shear rates (say, $\dot \gamma = 8~\si{\per \second}$) with heat transfer in the direction towards the outer wall ($\Delta T > 0$).

Finally, we investigated the effect of eccentricity of the Couette cell on the overall heat transfer characteristics. The eccentricity was varied from $E=0$ to $E=0.6$ above which jamming of particles (flow arrest) occurs. We observed an increase in the Nusselt number $Nu$ with $E$ for the $\Delta T> 0$ BC, while an optimum at $E=0.4$ exists for the $\Delta T < 0$ BC. Decrease in $Nu$ at larger $E$ (for the $\Delta T < 0$ case) is due to enhanced particle migration arising from the combined effect of shear- and thermal-induced particle migration aiding each other and a large recirculation zone resulting in a diminished particle flux, hence diminished heat transfer across the system. In addition, the heat transfer enhancement factor ($Nu_\mathrm{sus}/Nu_\mathrm{fc}$) is maximum for a concentric Couette cell. Even though $Nu_\mathrm{sus}$ and $Nu_\mathrm{fc}$ increase with eccentricity, their ratio decreases because particles are entrained near the outer wall for the suspension flow at large $E$, whereas they are not present in clear FC-43 fluid, and reduce the overall enhancement brought by the high particle conductivity.

Ultimately, Fig.~\ref{fig:NuvsE} shows that the range of $Nu$ is quite narrow for the temperature BCs under consideration because heat transfer in the system is dictated by the overall properties of the suspension, as also shown in previous studies \citep{Metzger2013HeatDiffusion,Dbouk2018HeatSimulation}. The main take-away from in this study that we wish to highlight to the reader is the novel effect of the temperature difference directionality on the enhancement or diminution of particle migration, arising from the thermo-rheological migration fluxes (Section~\ref{sec:thermo-rheol}). This effect is most clearly observed by contrasting the curves in Fig.~\ref{fig::thermalPe} for the two BCs. In future work, it would be of interest to experimentally interrogate this phenomenon.

In future work, it would be of interest to also explore engineering applications of the TFM for computational modeling of heat transfer in dense suspensions, along the lines of our preliminary study of electronics cooling via microchannels \cite{Nagrani2021Two-FluidCooling}. 

\subsection*{Acknowledgements}
Acknowledgment is made to the donors of the American Chemical Society Petroleum Research Fund for partial support of this research under ACS PRF award \#57371-DNI9, in particular for supporting P.P.N.\ and I.C.C.\ during the initial development of the dense suspension TFM and the \texttt{twoFluidsNBSuspensionFoam} solver \cite{Municchi2019TwoFluidsNBSuspensionFoam}. A.M.M.\ and I.C.C.\ are also affiliated with the Center for Particulate Products and Processes (CP\textsuperscript{3}) at Purdue University.

\printcredits

\bibliographystyle{cas-model2-names.bst}
\bibliography{Mendeley.bib,other.bib}

\clearpage
\appendix
\section{Calibration of the parameters in \texorpdfstring{$K_h$}{Kh} for non-zero shear rate}
\label{cal:shear_app}

\begin{figure}
    \centering 
    \includegraphics[width=0.6\textwidth]{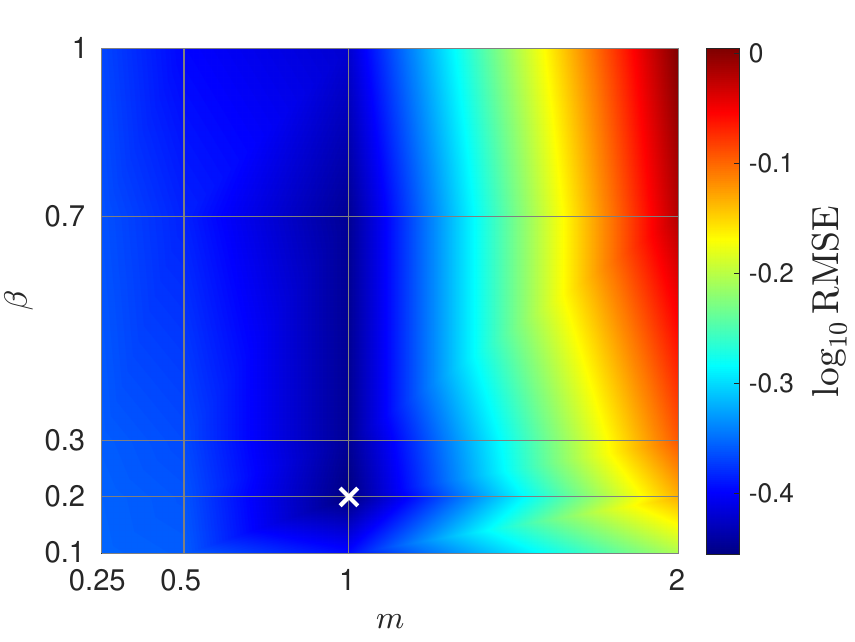}
    \caption{Root mean squared error ($\mathrm{RMSE}$) map between the transient temperature profiles of the heat-pulse experiment  \cite{Metzger2013HeatDiffusion} and the TFM simulations, computed for a range of parameter values $\beta \in [0.1,1.0]$ and $m\in[0.25,2]$ used in the closure for $K_h$ from Eq.~\eqref{eq::interPhaseHeatTransferCoeff}.}
    \label{fig:beta_m}
\end{figure}

Calibration of the closure relation in Eq.~\eqref{eq::interPhaseHeatTransferCoeff} is required to simulate coupled thermal-fluid suspension flows at non-zero shear rates. Specifically, the parameters $\beta$ and $m$ introduced in $K_h$ must be chosen to ensure that the simulated transient temperature decay profile at the inner cylinder wall matches that of the heat pulse experiment (recall Section~\ref{sec:Val_CC}). We considered $\beta \in [0.1,1.0]$ and $m\in[0.25,2]$ and calculated the root-mean-squared error (RMSE) between the TFM simulations and the experimental data as
\begin{equation}
    \mathrm{RMSE} = \sqrt{\frac{1}{n}\sum_{i=1}^{n} \left(T_{\mathrm{TFM},i}-T_{\mathrm{expt},i}\right)^2},
\end{equation}
where $n$ is total number of time points sampled. Here, $T_{\mathrm{TFM},i}$ is the temperature data generated from the TFM simulation by interpolating the transient temperature profile at the same time-points $t_i$ as the experimental measurement $T_{\mathrm{expt},i}$ (recall Fig.~\ref{fig::Val_CC}). As shown by the cross symbol in Fig.~\ref{fig:beta_m}, the RMSE is least for $\beta\approx0.2$ and $m\approx1$. Therefore, these are the parameter values used for the simulations in the main text.

\section{Calibration at zero shear rate: Lumped-parameter modeling of the cylinder}
\label{cal:no_shear}

Although not relevant to our study for \emph{shear}-induced migration, for completeness, here we discuss how the \emph{no shear} ($\dot\gamma = 0~\si{\per\second}$) curve is calibrated in Fig.~\ref{fig::Val_CC}. The inner and outer cylinders are held stationary for the entire simulation (during both the heating and cooling periods). For thermal BCs, although the outer wall is maintained at $293~\si{\kelvin}$ in the experiments, the gap is sufficiently large that we model the outer thermal BC as well insulated. For the inner cylinder, during the $5~\si{\second}$ of heating, we assume a constant wall temperature of $298~\si{\kelvin}$, which neglects the time it takes for the inner cylinder to heat up. For cooling, employing the idea from \cite[Section~2.2.2]{Metzger2013HeatDiffusion} used to estimate the thermal diffusivity of the suspension, we develop a modified time-decaying temperature gradient BC (for our simulations) at the inner wall. 

Specifically, consider the energy balance at the inner wall. Then, assuming a lumped capacitance model for the cylinder itself:
\begin{equation}
    \label{eq::energy_balance}
    M_\mathrm{cyl}C_P\frac{dT_\mathrm{wall}}{dt} = - 2\pi R_\mathrm{in} L \left[k_p \phi \frac{dT_p}{dr}  + k_f (1-\phi) \frac{dT_f}{dr}\right]_{r=R_\mathrm{in}},
\end{equation}
where $T_\mathrm{wall}$ is the inner cylinder wall temperature, $M_\mathrm{cyl}$, $C_P$ and $L$ are the mass, specific heat and length of the inner cylinder, respectively. To find the functional form of $dT_\mathrm{wall}/dt$ to use in Eq.~\eqref{eq::energy_balance} to obtain the sought after BC for simulations, we fit an exponential decay $T_\mathrm{wall}(t) -T_\mathrm{initial} = (T_\mathrm{max}-T_\mathrm{initial}) \exp(-A t)$ to the measured $T_\mathrm{wall}(t)$ in \cite{Metzger2013HeatDiffusion} (for the case of no shear), recalling that  $T_\mathrm{initial}=293~\si{\kelvin}$ and  $T_\mathrm{max}=298~\si{\kelvin}$. We obtained $A \approx 0.0129~\si{\per\second}$. Finding the expression for $dT_\mathrm{wall}/dt$ from this equation and substituting it into Eq.~\eqref{eq::energy_balance}, we solve for the radial temperature gradient at the inner wall, and obtain the following BC:
\begin{equation}
    \label{eq::zero_shear}
    \underbrace{\left[\phi \frac{dT_p}{dr}  + (1-\phi) \frac{dT_f}{dr}\right]_{r=R_\mathrm{in}}}_{\text{from simulation}} = \underbrace{B A \exp(-A t)}_{\text{from experiment}}.
\end{equation}

In this case, $k=k_p=k_f$, and hence it is included in $B = M_{\rm cyl} C_p (T_{\rm max}-T_{\rm initial})/ (2\pi R_\mathrm{in} L k)$, which lumps together all the unknown (unmeasured) inner cylinder physical quantities in this problem. Because there is no shear in this example, Eq.~\eqref{eq::interPhaseHeatTransferCoeff} reduces to $K_h = K_{h,0}$, which is obtained from the particle-based Ranz--Marshall Nusselt correlation. Thus, the only unknown in our model is $B$ in Eq.~\eqref{eq::zero_shear}, which is a constant that lumps together all the unknown inner cylinder properties. To find a suitable value for $B$, we performed TFM simulations, using Eq.~\eqref{eq::zero_shear} as the imposed BC, and matched the temperature decay profile predicted by the TFM to the experimentally measured one in \citep{Metzger2013HeatDiffusion}, to obtain $B \approx 10^5~\si{\kelvin \second \per \meter}$ (this is the simulation shown in Fig.~\ref{fig::Val_CC}).

\section{Mesh Independence Study}
\label{sec:appendix}

\begin{figure}
    \centering
    \includegraphics[width=0.6\linewidth]{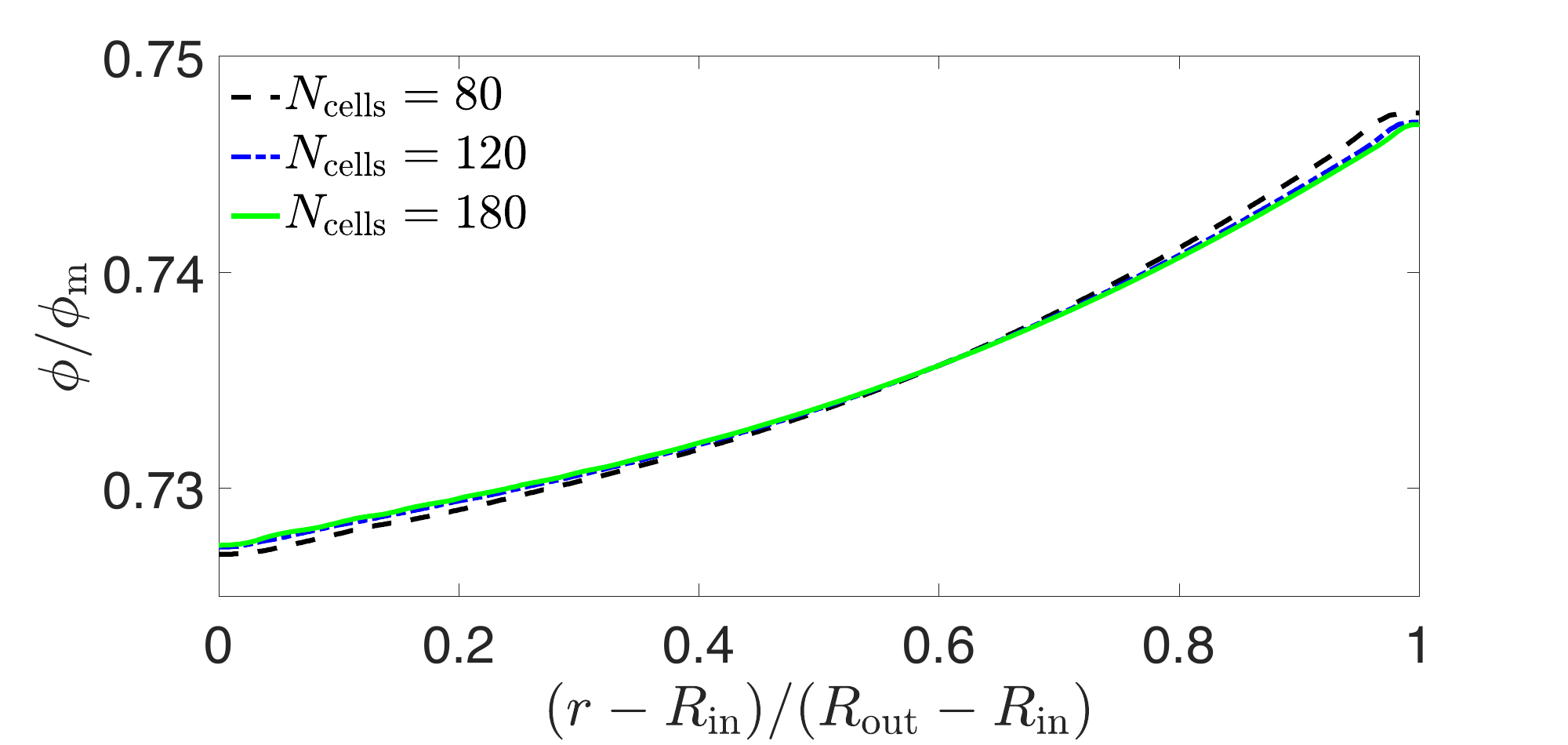}
    \caption{Mesh independent study for the concentric Couette cell. Here, $N_{\rm cells}$ is the number of mesh cells in the radial direction across the gap. The grid arrangement is shown in Fig.~\ref{fig::geoMesh}.}
    \label{fig::grid_Independent}
\end{figure}

In this appendix, we justify the mesh used to produce the simulation results in the main text. To be able to trust our conclusions, we must verify 
the simulations are independent of the mesh resolution (in addition to the calibration/validation against experiments performed in Sections~\ref{sec:Val_CC} and \ref{sec:Val_EC}). We consider the concentric Couette cell geometry (Fig.~\ref{fig::geoMesh}) filled with $d_p=675~\si{\micro\meter}$ BN particles suspended at $\phi_b=50\%$ in FC-43 fluid (refer to Table~\ref{tab::sus_props_BN_FC} for the remaining properties). A temperature difference of $\Delta T = 30~\si{\kelvin}$ is set across the Couette cell, in which the inner cylinder is rotated at a shear rate of $\dot{\gamma} = 3~\si{\per\second}$. As seen from Fig.~\ref{fig::grid_Independent}, $N_{\rm cells}=120$ along the radial direction accurately captures the particle volume fraction distribution, providing us with the computationally ``optimal'' mesh resolution.

\end{document}